\documentclass[iop,apj]{emulateapj}
\usepackage{apjfonts}
\usepackage{graphicx,float,wrapfig}
\usepackage{multirow, threeparttable, color}
\usepackage{hyperref}
\usepackage{rotating}
\usepackage{ulem}
\hypersetup{
    colorlinks,
    citecolor=blue,
    filecolor=blue,
    linkcolor=blue,
    urlcolor=blue
}

\makeatletter
\newcommand*{\rom}[1]{\expandafter\@slowromancap\romannumeral #1@}
\makeatother 

\newcommand{\fesc}{\ifmmode{f_{\rm esc}}\else{$f_{\rm esc}$}\fi}
\newcommand{\fescs}{\ifmmode{f_{\rm esc}^\star}\else{$f_{\rm esc}^\star$}\fi}
\newcommand{\kms}{\ifmmode{{\;\rm km~s^{-1}}}\else{km~s$^{-1}$}\fi}
\newcommand{\fgas}{\ifmmode{{f_{\rm gas}}}\else{$f_{\rm gas}$}\fi}
\newcommand{\cubecm}{\ifmmode{{\rm cm^{-3}}}\else{cm$^{-3}$}\fi}
\newcommand{\ztwo}{\ifmmode{{\rm [Z_2/H]}}\else{[Z$_2$/H]}\fi}
\newcommand{\zthree}{\ifmmode{{\rm [Z_3/H]}}\else{[Z$_3$/H]}\fi}
\newcommand{\lsim}{\lower0.3em\hbox{$\,\buildrel <\over\sim\,$}}
\newcommand{\gsim}{\lower0.3em\hbox{$\,\buildrel >\over\sim\,$}}

\newcommand{\sfr}{\ifmmode{\textrm{M}_\odot \,\textrm{yr}^{-1} \,\textrm{Mpc}^{-3}}\else{M$_\odot$ yr$^{-1}$ Mpc$^{-3}$}\fi}
\newcommand{\hsfr}{\ifmmode{\;\textrm{M}_\odot\, \textrm{yr}^{-1}}\else{M$_\odot$ yr$^{-1}$}\fi}

\newcommand{\eavg}{\ifmmode{\langle E_\gamma \rangle}\else{$\langle E_\gamma \rangle$}\fi}

\newcommand{\Ms}{\ifmmode{M_\odot}\else{$M_\odot$}\fi}
\newcommand{\vrms}{\ifmmode{v_{\rm rms}}\else{$v_{\rm rms}$}\fi}

\newcommand{\tvir}{\ifmmode{T_{\rm{vir}}}\else{$T_{\rm{vir}}$}\fi}
\newcommand{\mvir}{\ifmmode{M_{\rm{vir}}}\else{$M_{\rm{vir}}$}\fi}
\newcommand{\rvir}{\ifmmode{r_{\rm{vir}}}\else{$r_{\rm{vir}}$}\fi}

\newcommand{\jj}{\ifmmode{J_{21}}\else{$J_{21}$}\fi}
\newcommand{\flw}{\ifmmode{F_{LW}}\else{$F_{LW}$}\fi}
\newcommand{\kph}{\ifmmode{k_{\rm ph}}\else{$k_{\rm ph}$}\fi}

\newcommand{\zsun}{\ifmmode{\rm\,Z_\odot}\else{$\rm\,Z_\odot$}\fi}

\newcommand{\hi}{H {\sc i}}
\newcommand{\hii}{H {\sc ii}}

\newcommand{\nhi}{\ifmmode{N_{\rm HI}}\else{$N_{\rm HI}$}\fi}

\newcommand{\mh}{M_h}

\begin{document}

\shorttitle{The role of the smallest galaxies in reionization}
\shortauthors{Norman et al.}
\journalinfo{Submitted to the Astrophysical Journal}
\submitted{Draft version \today}

\title{Fully Coupled Simulation of Cosmic Reionization. III. Stochastic Early Reionization by the Smallest Galaxies}
\author{ 
  Michael L. Norman\altaffilmark{1,2}, 
  Pengfei Chen\altaffilmark{1},
  John H. Wise\altaffilmark{3},
  and Hao Xu\altaffilmark{2}
}

\affil{$^{1}${CASS,
  University of California, San Diego, 9500 Gilman Drive, La Jolla, CA
  92093; 
  \href{mailto:pec008@ucsd.edu}{pec008@ucsd.edu}, 
  \href{mailto:mlnorman@ucsd.edu}{mlnorman@ucsd.edu}}}
  
  \affil{$^{2}${San Diego Supercomputer Center, 9500 Gilman Drive, La Jolla, CA
  92093; \href{mailto:hxu@ucsd.edu}{hxu@ucsd.edu}}}

\affil{$^{3}${Center for Relativistic Astrophysics, School of
  Physics, Georgia Institute of Technology, 837 State Street, Atlanta,
  GA 30332; \href{mailto:jwise@gatech.edu}{jwise@gatech.edu}}}


\begin{abstract}

Previously we identified a new class of early galaxy that we estimate contributes up to 30\% of the ionizing photons responsible for reionization. These are low mass halos in the range $M_h =10^{6.5}-10^{8} M_{\odot}$ that have been chemically enriched by supernova ejecta from prior Pop III star formation. Despite their low star formation rates, these Metal Cooling halos (MCs) are significant sources of ionizing radiation, especially at the onset of reionization, due to their high number density and ionizing escape fractions. Here we present a fully-coupled radiation hydrodynamic simulation of reionization that includes these MCs as well the more massive hydrogen atomic line cooling halos. Our method is novel: we perform halo finding inline with the radiation hydrodynamical simulation, and assign escaping ionizing fluxes to halos using a probability distribution function (PDF) measured from the galaxy-resolving {\it Renaissance Simulations}. The PDF captures the mass dependence of the ionizing escape fraction as well as the probability that a halo is actively forming stars.  With MCs, reionization starts earlier than if only halos of $10^8 M_{\odot}$ and above are included, however the redshift when reionization completes is only marginally affected as this is driven by more massive galaxies.  Because star formation is intermittent in MCs, the earliest phase of reionization exhibits a stochastic nature, with small \hii~ regions forming and recombining. Only later, once halos of mass $\sim 10^9 M_{\odot}$ and above begin to dominate the ionizing emissivity, does reionization proceed smoothly in the usual manner deduced from previous studies. This occurs at $z\approx 10$ in our simulation.  
\end{abstract}


\section{Introduction}
The relative contributions to reionization from halos in different mass ranges are still not clear. A useful taxonomy for discussion we follow here was introduced by \cite{Iliev07}. The halos hosting early galaxies could be divided into three categories according to their mass. The first are minihalos (MHs, $\mh <10^8M_\odot$), which host the formation of Population III stars but otherwise are not thought to be efficient star formers due to their low virial temperatures and low $H_2$ cooling efficiency. The second are low-mass atomic-cooling halos (LMACHs, $10^8M_\odot<\mh<10^9M_\odot$), which have virial temperatures just above the threshold to excite H atomic line cooling and form stars inefficiently. The third are high-mass atomic cooling halos (HMACHs, $\mh>10^9M_\odot$), which cool and form stars more efficiently than the LMACHs. 

Some work has been done to study the role of LMACHs and MHs in reionization \citep{Iliev07, Choudhury08, Shapiro12, Iliev12, Ahn12, Wyithe13} which show that with the smallest galaxies included reionization begins earlier and the intergalactic electron-scattering optical depth $\tau_{es}$ is boosted. However these authors find the late phase of reionization is still dominated by HMACHs and the overlap redshift $z_{ov}$ is not significantly affected. In these studies the galaxy properties are not simulated directly, but rather assumed using simple parameterized models which directly relate a halo's mass to its ionizing emissivity. For the smallest galaxies this relation is exceedingly uncertain due to a variety of complex physical processes. For example, the formation of the smallest galaxies is possibly suppressed due to the large Lyman-Werner background which photodissociates the primary coolant $H_2$ \citep{Ahn12}, and due to supernova feedback which depletes the halo of gas \citep{Wyithe13}. Some simulations are used to predict the signatures of reionization on the high redshift 21cm background, and to discuss how 21cm observations could help to distinguish the relative contributions of galaxies of different masses to reionization \citep{Shapiro12, Iliev12}.

However, recently \cite{Wise14} have shown using AMR radiation hydrodynamic simulations that minihalos which have been chemically enriched by supernova ejecta from prior Pop III star formation can cool and form stars, and moreover significantly contribute to the overall ionizing photon budget of reionization. We refer to this new class of halos as metal-line cooling halos, or MCs. Follow-on simulations (the {\it Renaissance Simulations}) in much larger volumes by \cite{Xu16} provide the star formation rates (SFR), intermittency, and ionizing escape fractions in the MCs, LMACHs, and HMACHs with extremely high resolution and good statistics. Using these results as our input, we revisit the problem: what is the role of the lowest mass halos in the reionization process? The main improvement of our work compared with previous work is that the simulations shown here are fully coupled cosmological radiation hydrodynamic simulations, with a time-dependent treatment of the ionization kinetics, and emissivities assigned to the source halos dynamically, considering the intermittency of the contribution from MCs. 

We find that because star formation is intermittent in MCs, the earliest phases of reionization exhibits a stochastic nature, with small \hii~ regions forming and recombining. Only later, after the characteristic halo mass scale has reached $\sim 10^9\Ms$, does reionization proceed smoothly in the usual manner deduced from earlier studies. 
Adopting concordance cosmological parameters and only using the galaxy properties from the {\it  Renaissance Simulations}, our $1152^3$ simulation in a 14.4 comoving Mpc box begins reionizing at $z=20$, is $10\%$ ionized at $z=10$, and fully ionizes at $ z=7.1$.  

This paper is organized as follows. We summarize the relevant results from \cite{Xu16} in Section 2. The description of computational method and inputs to the simulations is provided in Section 3. We show results in Section 4 and offer discussion and conclusions in Section 5.

\section{Ionizing Photons from the Smallest Galaxies}

\begin{figure*}
\centering \includegraphics[width=1.8\columnwidth]{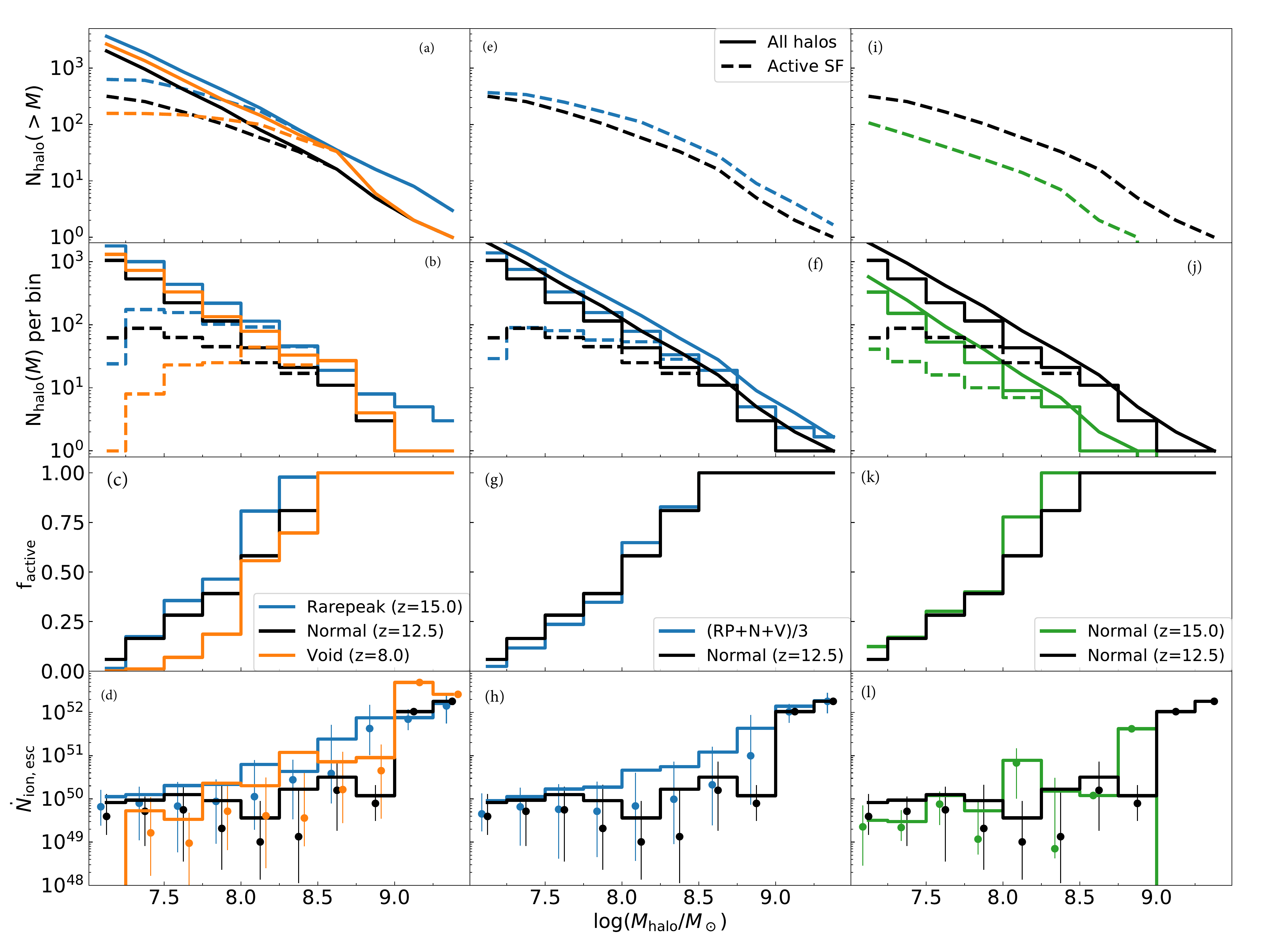}
\caption{Distribution of halo counts and escaped ionizing photons in the {\it Renaissance Simulations} from which our $\mh - \dot{N}_{ion,esc}$ model is derived. First row: cumulative halo counts for all halos (solid lines) and halos actively forming stars (dashed lines). Second row: number of halos in 0.25 dex bins for all halos and those actively forming stars. Third row: fraction of halos in bins actively forming stars. Fourth row: mean number of escaping ionizing photons per bin (solid lines), median value (points), and 1-sigma error bars (thin vertical lines). First column: statistics for 3 Renaissance Simulations at their stopping redshifts: Rarepeak at z=15, Normal at z=12.5, and Void at z=8. Second column: comparison with Normal simulation statistics at z=12.5 with one third of the total sample at their stopping redshifts (see text for justification). Third column: statistics for the Normal simulation at z=15 and z=12.5. The bottom row shows that the mean number of escaping ionizing photons per halo in the Normal simulation is roughly constant at $\sim 10^{50}$ s$^{-1}$ over the mass range $10^7 - 10^{8.5} \Ms$.  } 
\label{fig:halo-counts}
\end{figure*}

Unlike the previous papers in this series \citep{So14,Norman13}, where ionizing emissivites were calculated from a simple star formation/feedback recipe incorporated in the simulation itself, here we employ results from much higher resolution simulations which calculate the escaping ionizing photons of high redshift galaxies directly. How this is done is described in Secs. \ref{halofinding} and \ref{input}. This approach enormously relaxes the spatial resolution requirement on the global reionization simulation and takes advantage of more precise simulation results. Uniform grids may be employed for the reionization simulation, however they must have sufficient mass and spatial resolution to accurately capture the halo population of importance. In addition, we are able to use moment methods for the radiation transport, which do not get bogged down as reionization completes as some ray tracing methods do \citep{Norman13}. 

We draw on the results of  \cite{Xu16} who performed three high-resolution AMR simulations of regions of different over-densities in order to study the abundance, environmental dependence, and escape fraction $f_{esc}$ of the smallest galaxies during reionization. The so-called {\it Renaissance Simulations} include both Pop II and Pop III star formation and their radiative, mechanical, and chemical feedback. The three simulations are named Void (V), Normal (N), and Rarepeak (RP), as they respectively simulate underdense, average, and overdense regions within a larger cosmological density field. The simulations were run to their stopping redshifts of z=8, 12.5, and 15, respectively, where they each produce roughly 3000 MCs, several hundred LMACHs, and a handful of HMACHs  [\citet{Xu16}, Table 1].  

\begin{figure}
\includegraphics[width=1.0\columnwidth]{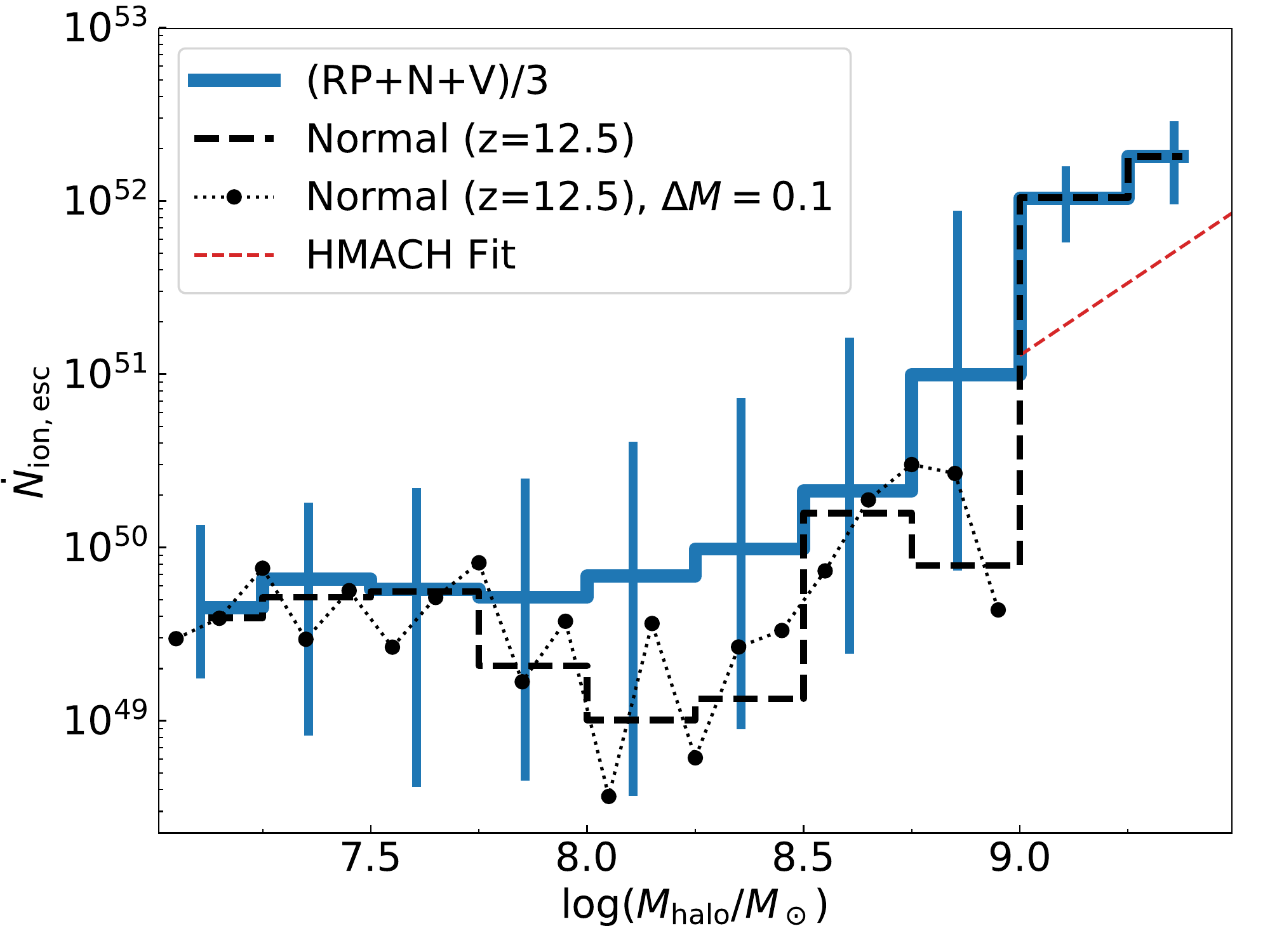}
\caption{Effect of mass binning on the escaping ionizing fluxes in the Normal simulation. Black dotted and dashed lines show median emissivities for the Normal simulation at z=12.5 in bins of 0.1 and 0.25 dex, respectively. The blue histogram is the distribution for the combined sample (RP+N+V)/3 with $1\sigma$ error bars. We see that Table 1, which uses 0.1 dex mass bins, is consistent with the 0.25 dex binning as well as the larger sample within their $1-\sigma$ errors.} 
\label{fig:nion-compare}
\end{figure}

\begin{figure}
\includegraphics[width=1.0\columnwidth]{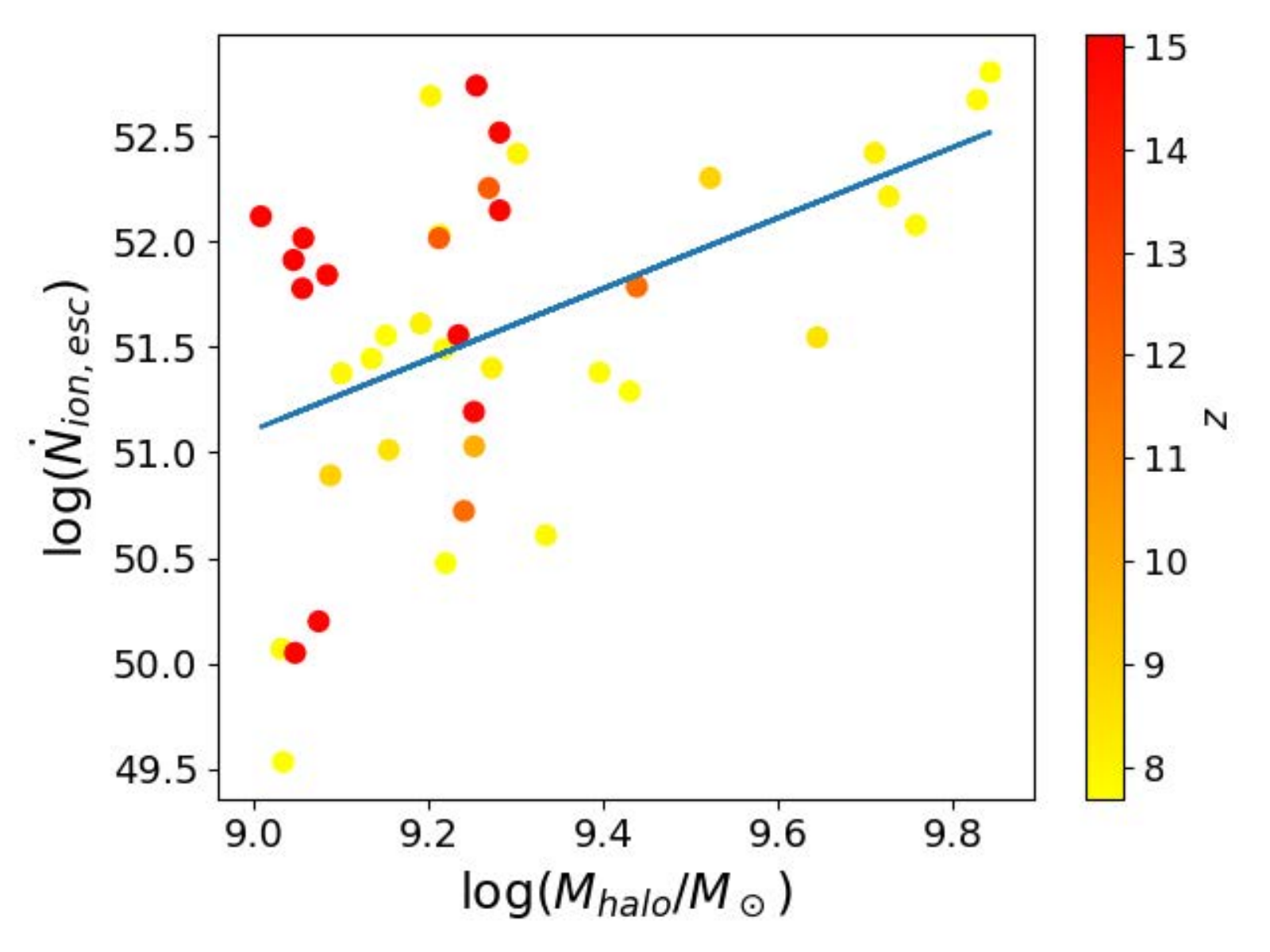}
\caption{Escaping ionizing photons for the combined HMACH sample from all Renaissance Simulations color-coded by redshift (points) and a fit (solid line) used in the reionization simulations presented in this paper.}
\label{fig:hmach_fit}
\end{figure}

The halo mass functions are shown in Fig. \ref{fig:halo-counts}a. Solid lines plot all the halos, while dashed lines plot halos actively forming stars (halos with stars younger the 20 Myr).  Fig. \ref{fig:halo-counts}b plots differential halo counts in 0.25 dex mass bins, while Fig. \ref{fig:halo-counts}c plots the fraction of halos actively forming stars within those same mass bins. The fraction of halos actively forming stars, and hence contributing to reionization, rises with halo mass, reaching unity around  $\mh \approx 10^{8.5} \Ms$ in all 3 simulations.  Fig. \ref{fig:halo-counts}d plots the mean escaping ionizing photon emissivity per bin (solid lines), which shows a relatively flat distribution for the Normal simulation below $10^{8.5} \Ms$, whereas as the Void and Rarepeak simulations shows an increasing distribution. The median escaping ionizing photon emissivity per bin is plotted as points, with vertical lines depicting the 1$\sigma$ standard deviation intervals. Figs. \ref{fig:halo-counts}e-h plot the same information as  Fig. \ref{fig:halo-counts}a-d for the Normal simulation alone, and compares it with an average of the Rarepeak, Normal, and Void simulations at their stopping redshifts desginated (RP+N+V)/3 (see below for justification for this averaging). We do this to investigate whether the Normal simulation is representative of the larger sample. The ionizing emissivity distributions are analyzed in further detail below. Fig. \ref{fig:halo-counts}g shows that the fraction of halos actively forming stars is indeed representative of the larger sample. Figs. \ref{fig:halo-counts}i-l plot the same information as  Fig. \ref{fig:halo-counts}a-d for alone at two redshifts: z=15 and z=12.5. We see an overall increase in the number of halos with decreasing redshift, but only minor evolution in the fraction of active halos per mass bin. 

We adopt the properties of the galaxies in the Normal simulation at z=12.5 as the basis for our halo ionizing emissivity model.  There are two reasons for this. First, we simulate an average region of the universe, and therefore the Normal simulation is most appropriate. Also, our simulated volume is too small to contain the statistical extremes that the Rarepeak and Void simulations represent. Second, we are most interested in the contribution of the smallest galaxies to the early stages of reionization $20 \leq z \leq 10$. As we will show, over this redshift range the MCs and LMACHs dominate the HMACHs. Fig. \ref{fig:halo-counts}b shows that we have 50-60 active halos per bin in the MC mass range, 20-30 active halos per bin in the lower LMACH mass range   $10^8 \leq \mh/\Ms \leq 10^{8.5}$, and 4-10 active halos per bin in the upper LMACH mass range   $10^{8.5} \leq \mh/\Ms \leq 10^{9}$. Thus, our statistical coverage is reasonably good over the mass range  $10^7 \leq \mh/\Ms \leq 10^{8.5}$, i.e., halos that form stars intermittently, becoming somewhat noisy in the mass range  $10^{8.5} \leq \mh/\Ms \leq 10^{9}$. 

An important finding of \cite{Xu16} is that the galaxies from the three Renaissance Simulations at their different stopping redshifts form a homogeneous single population in which the halo mass is the dominant parameter determining an individual galaxy's properties. This is best illustrated by Fig. 15 of that paper, which overlays scatter plots of stellar mass, star formation rate, stellar baryon fraction, and ionizing escape fraction, all versus halo virial mass. The point clouds from the three simulations at their different stopping redshifts overlap one another with no obvious differences. Their similarity can also be seen in the evolution of the stellar mass function in the three simulations, which proceed in qualitatively the same way, but just offset in redshift. These similarities justifies combining our three halo samples into a single larger halo sample for the purposes of examining how reprentative is the Normal galaxy sample relative to the combined sample. Fig. \ref{fig:halo-counts}e-h compares the z=12.5 Normal halo statistics with the combined sample. The median $\dot{N}_{\rm ion,esc}$ value in each individual region (panel d) is not a monotonic function of halo mass, but when averaged over all three regions (panel h), the dependence on halo mass becomes more smooth, albeit with additional scatter.  It is evident from this comparison that the z=12.5 Normal halos are indeed representative of the larger sample, drawn from different environments and redshifts over the mass range $10^7 \leq \mh/\Ms \leq 10^{8.5}$, but somewhat under-represents the massive end of the LMACHs $10^{8.5} \leq \mh/\Ms \leq 10^{9}$. 

From the analysis presented in Figure \ref{fig:halo-counts}, there is no clear trend with respect to redshift or environment, as shown in \citet{Xu16}.  The MCs and LMACHs form stars intermittently on the order of 20-40 Myr (see Figure \ref{fig:fesc-lmach} in the Discussion), and thus at a given time, they can either be quiescent or actively forming stars, resulting in a large scatter in their escaping ionizing luminosities, which can be seen in the bottom row of Figure \ref{fig:halo-counts}.  Although the Normal simulation is statistically consistent with the (RP+N+V)/3 sample, the $10^{8}$, $10^{8.25}$, and $10^{8.75}~\Ms$ mass bins differ by an order of magnitude.  This variation in two mass bins ($10^8$ and $10^{8.75}~\Ms$) is also apparent in the Normal simulation at z=15 and z=12.5.  If we were to have adopted the total (RP+N+V)/3 sample in this study, the total emissivity would have been slightly boosted in the initial stages of reionization ($20 \leq z \leq 10$) when the ionized volume fraction is only a few percent.  Thus we do not expect these differences to have a large impact on the reionization history and completion redshift, and we defer a further discussion of this impact to Section \ref{subsec:limitations}.

Our halo emissivity model for MCs and LMACHs is constructed from the data presented in Fig. 19d of \cite{Xu16}, which displays a 2D histogram of ionizing emissivities for halos within a given mass bin from the Normal simulation at z=12.5. This is the same data that were used for Fig. \ref{fig:halo-counts}d.  Because of the considerable amount of scatter within each mass bin, we probabilistically assign halos in the mass range $10^7\Ms<\mh<10^9\Ms$ an ionizing emissivity $\dot{N}_{ion,esc}$ according to a 2D lookup table, reproduced in Table 1. This table is constructed as follows: for each mass bin, here chosen to be 0.1 dex wide, emissivities are set to zero for probabilities $0\leq p \leq 1- f_{active}$, where $f_{active}$ is the fraction of halos actively forming stars as shown in Fig. 1. The non-zero values within a mass bin are set according to the 1D PDF of ionizing emissivities shown in Fig. 19d of  \cite{Xu16}. By counting the number of repeated values within a bin, one can see that the statistical coverage for halos in the mass range $10^7\Ms<\mh<10^{8.5}\Ms$ is reasonably good. However our statistical coverage over the mass range $10^{8.6}\Ms<\mh<10^9\Ms$ is somewhat poorer, with only 15 unique halos. It is therefore important to ask how representative is Table 1 of the ionizing emissivity distribution functions from the other Renaissance Simulations at different redshifts. 

This is addressed in Fig. \ref{fig:nion-compare}. Black dotted and dashed lines show median emissivities for the Normal simulation at z=12.5 in mass bins of width 0.1 and 0.25 dex, respectively. The blue histogram is the median distribution for the combined sample (RP+N+V)/3 with $1\sigma$ error bars superposed. We see that Table 1, which uses 0.1 dex mass bins, is consistent with the 0.25 dex binning as well as the larger sample at the 1$\sigma$ level, with the exception of the very last bin at log($\mh/\Ms = 8.9)$, which is more than 1$\sigma$ low. This is a statistical fluctuation due to the fact that we only have 2 halos in that mass bin. The effect of this low value will be discussed in Section \ref{sec:discussion} after we have presented our results. 

For HMACHs, we assign emissivities according to a fit obtained from a combined sample of all halos exceeding $10^9 \Ms$ from all three simulations at all redshifts (cf. Fig. \ref{fig:hmach_fit}):
\begin{equation}
  \label{eqn:emis9}
\rm{log}_{10}(\dot{N}_{\rm ion,esc}) = 36.033+1.675\times \rm{log}_{10}(\mh/\Ms) ~sec^{-1}.
\end{equation}

\noindent 
The fit is included in Fig. \ref{fig:nion-compare} as a red dashed line.

\begin{table*}
\centering
\caption{$M_{\rm halo}$ dependent escaped ionizing photons rate $\dot{N}_{\rm ion,esc}$ used in all simulations in Table 2 for halos in the mass range  $10^7 \leq \mh/\Ms \leq 10^9$. See text for a description how this table was constructed. $\mh$ is in units of $\Ms$; $\dot{N}_{\rm ion,esc}$ is in units of sec$^{-1}$.}
\label{tab_hao} 
\begin{tabular}{ccccccccccccccccccccc}
\hline\hline
& $10^{7.0}$  & $10^{7.1}$  & $10^{7.2}$  & $10^{7.3}$  & $10^{7.4}$  & $10^{7.5}$  & $10^{7.6}$  & $10^{7.7}$  & $10^{7.8}$  & $10^{7.9}$  \\
\hline
0.00  &0.00 &0.00 &0.00 &0.00 &0.00 &0.00 &0.00 &0.00 &0.00 &0.00 \\
\hline
...
&... &... &... &... &... &... &... &... &... &... \\
\hline
0.76  &0.00 &0.00 &0.00 &0.00 &0.00 &0.00 &0.00 &0.00 &0.00 &0.00 \\
\hline
0.78  &0.00 &0.00 &0.00 &0.00 &0.00 &0.00 &0.00 &0.00 &0.00 &$1.20\times 10^{46}$ \\
\hline
0.80  &0.00 &0.00 &0.00 &0.00 &0.00 &0.00 &0.00 &0.00 &0.00 &$3.08\times 10^{48}$ \\
\hline
0.82  &0.00 &0.00 &0.00 &0.00 &0.00 &0.00 &0.00 &0.00 &$9.60\times 10^{47}$ &$3.37\times 10^{48}$ \\
\hline
0.84  &0.00 &0.00 &0.00 &0.00 &0.00 &0.00 &0.00 &0.00 &$3.67\times 10^{48}$ &$5.19\times 10^{48}$ \\
\hline
0.86  &0.00 &0.00 &0.00 &0.00 &0.00 &0.00 &0.00 &0.00 &$7.20\times 10^{48}$ &$2.34\times 10^{49}$ \\
\hline
0.88  &0.00 &0.00 &0.00 &0.00 &0.00 &0.00 &0.00 &$1.90\times 10^{48}$ &$1.56\times 10^{49}$ &$5.04\times 10^{49}$ \\
\hline
0.90  &0.00 &0.00 &0.00 &$1.44\times 10^{48}$ &$6.40\times 10^{48}$ &$1.12\times 10^{46}$ &$4.42\times 10^{46}$ &$8.67\times 10^{48}$ &$1.66\times 10^{49}$ &$8.86\times 10^{49}$ \\
\hline
0.92  &0.00 &0.00 &0.00 &$1.68\times 10^{49}$ &$1.44\times 10^{49}$ &$1.81\times 10^{49}$ &$1.39\times 10^{49}$ &$5.68\times 10^{49}$ &$2.35\times 10^{49}$ &$1.39\times 10^{50}$ \\
\hline
0.94  &0.00 &0.00 &$5.50\times 10^{48}$ &$5.41\times 10^{49}$ &$5.18\times 10^{49}$ &$4.11\times 10^{49}$ &$4.64\times 10^{49}$ &$9.74\times 10^{49}$ &$2.87\times 10^{49}$ &$1.81\times 10^{50}$ \\
\hline
0.96  &0.00 &$5.35\times 10^{49}$ &$3.81\times 10^{49}$ &$7.95\times 10^{49}$ &$1.97\times 10^{50}$ &$7.44\times 10^{49}$ &$1.12\times 10^{50}$ &$2.38\times 10^{50}$ &$1.02\times 10^{50}$ &$1.88\times 10^{50}$ \\
\hline
0.98  &$1.82\times 10^{50}$ &$3.35\times 10^{50}$ &$1.69\times 10^{50}$ &$2.43\times 10^{51}$ &$5.21\times 10^{50}$ &$3.49\times 10^{50}$ &$3.00\times 10^{50}$ &$3.71\times 10^{50}$ &$1.67\times 10^{50}$ &$7.50\times 10^{50}$ \\
\hline
\hline
& $10^{8.0}$  & $10^{8.1}$  & $10^{8.2}$  & $10^{8.3}$  & $10^{8.4}$  & $10^{8.5}$  & $10^{8.6}$  & $10^{8.7}$  & $10^{8.8}$  & $10^{8.9}$  \\
\hline
0.00  &0.00 &0.00 &0.00 &0.00 &0.00 &0.00 &$9.30\times 10^{48}$ &$8.15\times 10^{49}$ &$4.91\times 10^{49}$ &$1.12\times 10^{50}$ \\
\hline
0.02  &0.00 &0.00 &0.00 &0.00 &0.00 &0.00 &$9.30\times 10^{48}$ &$8.15\times 10^{49}$ &$4.91\times 10^{49}$ &$1.12\times 10^{50}$ \\
\hline
0.04  &0.00 &0.00 &0.00 &0.00 &0.00 &0.00 &$9.30\times 10^{48}$ &$8.15\times 10^{49}$ &$4.91\times 10^{49}$ &$1.12\times 10^{50}$ \\
\hline
0.06  &0.00 &0.00 &0.00 &0.00 &0.00 &0.00 &$9.30\times 10^{48}$ &$8.15\times 10^{49}$ &$4.91\times 10^{49}$ &$1.12\times 10^{50}$ \\
\hline
0.08  &0.00 &0.00 &0.00 &0.00 &0.00 &0.00 &$9.30\times 10^{48}$ &$8.15\times 10^{49}$ &$4.91\times 10^{49}$ &$1.12\times 10^{50}$ \\
\hline
0.10  &0.00 &0.00 &0.00 &0.00 &0.00 &0.00 &$9.30\times 10^{48}$ &$8.15\times 10^{49}$ &$4.91\times 10^{49}$ &$1.12\times 10^{50}$ \\
\hline
0.12  &0.00 &0.00 &0.00 &0.00 &0.00 &0.00 &$9.30\times 10^{48}$ &$8.15\times 10^{49}$ &$4.91\times 10^{49}$ &$1.12\times 10^{50}$ \\
\hline
0.14  &0.00 &0.00 &0.00 &0.00 &0.00 &0.00 &$9.30\times 10^{48}$ &$8.15\times 10^{49}$ &$4.91\times 10^{49}$ &$1.12\times 10^{50}$ \\
\hline
0.16  &0.00 &0.00 &0.00 &0.00 &0.00 &0.00 &$9.30\times 10^{48}$ &$1.64\times 10^{50}$ &$4.91\times 10^{49}$ &$1.12\times 10^{50}$ \\
\hline
0.18  &0.00 &0.00 &0.00 &0.00 &0.00 &0.00 &$9.30\times 10^{48}$ &$1.64\times 10^{50}$ &$4.91\times 10^{49}$ &$1.12\times 10^{50}$ \\
\hline
0.20  &0.00 &0.00 &0.00 &0.00 &$1.45\times 10^{47}$ &0.00 &$9.30\times 10^{48}$ &$1.64\times 10^{50}$ &$4.91\times 10^{49}$ &$1.12\times 10^{50}$ \\
\hline
0.22  &0.00 &0.00 &0.00 &0.00 &$1.45\times 10^{47}$ &0.00 &$9.30\times 10^{48}$ &$1.64\times 10^{50}$ &$4.91\times 10^{49}$ &$1.12\times 10^{50}$ \\
\hline
0.24  &0.00 &0.00 &0.00 &0.00 &$1.45\times 10^{47}$ &0.00 &$9.83\times 10^{48}$ &$1.64\times 10^{50}$ &$4.91\times 10^{49}$ &$1.12\times 10^{50}$ \\
\hline
0.26  &0.00 &0.00 &0.00 &0.00 &$1.45\times 10^{47}$ &0.00 &$9.83\times 10^{48}$ &$1.64\times 10^{50}$ &$4.91\times 10^{49}$ &$1.12\times 10^{50}$ \\
\hline
0.28  &0.00 &0.00 &0.00 &0.00 &$1.45\times 10^{47}$ &0.00 &$9.83\times 10^{48}$ &$1.64\times 10^{50}$ &$4.91\times 10^{49}$ &$1.12\times 10^{50}$ \\
\hline
0.30  &0.00 &0.00 &0.00 &0.00 &$1.45\times 10^{47}$ &0.00 &$9.83\times 10^{48}$ &$1.64\times 10^{50}$ &$4.91\times 10^{49}$ &$1.12\times 10^{50}$ \\
\hline
0.32  &0.00 &0.00 &0.00 &0.00 &$1.45\times 10^{47}$ &0.00 &$9.83\times 10^{48}$ &$3.92\times 10^{50}$ &$6.88\times 10^{50}$ &$1.12\times 10^{50}$ \\
\hline
0.34  &0.00 &0.00 &0.00 &0.00 &$1.45\times 10^{47}$ &0.00 &$9.83\times 10^{48}$ &$3.92\times 10^{50}$ &$6.88\times 10^{50}$ &$1.12\times 10^{50}$ \\
\hline
0.36  &0.00 &0.00 &0.00 &0.00 &$1.45\times 10^{47}$ &0.00 &$9.83\times 10^{48}$ &$3.92\times 10^{50}$ &$6.88\times 10^{50}$ &$1.12\times 10^{50}$ \\
\hline
0.38  &0.00 &0.00 &0.00 &0.00 &$1.45\times 10^{47}$ &0.00 &$9.83\times 10^{48}$ &$3.92\times 10^{50}$ &$6.88\times 10^{50}$ &$1.12\times 10^{50}$ \\
\hline
0.40  &0.00 &0.00 &0.00 &0.00 &$6.38\times 10^{48}$ &0.00 &$9.83\times 10^{48}$ &$3.92\times 10^{50}$ &$6.88\times 10^{50}$ &$1.12\times 10^{50}$ \\
\hline
0.42  &0.00 &0.00 &0.00 &0.00 &$6.38\times 10^{48}$ &0.00 &$9.83\times 10^{48}$ &$3.92\times 10^{50}$ &$6.88\times 10^{50}$ &$1.12\times 10^{50}$ \\
\hline
0.44  &0.00 &0.00 &0.00 &0.00 &$6.38\times 10^{48}$ &0.00 &$9.83\times 10^{48}$ &$3.92\times 10^{50}$ &$6.88\times 10^{50}$ &$1.12\times 10^{50}$ \\
\hline
0.46  &0.00 &0.00 &0.00 &0.00 &$6.38\times 10^{48}$ &0.00 &$9.83\times 10^{48}$ &$3.92\times 10^{50}$ &$6.88\times 10^{50}$ &$1.12\times 10^{50}$ \\
\hline
0.48  &0.00 &0.00 &0.00 &0.00 &$6.38\times 10^{48}$ &0.00 &$9.83\times 10^{48}$ &$3.92\times 10^{50}$ &$6.88\times 10^{50}$ &$1.12\times 10^{50}$ \\
\hline
0.50  &0.00 &0.00 &0.00 &0.00 &$6.38\times 10^{48}$ &0.00 &$1.16\times 10^{49}$ &$5.39\times 10^{50}$ &$6.88\times 10^{50}$ &$1.54\times 10^{51}$ \\
\hline
0.52  &0.00 &0.00 &0.00 &0.00 &$6.38\times 10^{48}$ &0.00 &$1.16\times 10^{49}$ &$5.39\times 10^{50}$ &$6.88\times 10^{50}$ &$1.54\times 10^{51}$ \\
\hline
0.54  &0.00 &0.00 &0.00 &0.00 &$6.38\times 10^{48}$ &0.00 &$1.16\times 10^{49}$ &$5.39\times 10^{50}$ &$6.88\times 10^{50}$ &$1.54\times 10^{51}$ \\
\hline
0.56  &0.00 &0.00 &0.00 &0.00 &$6.38\times 10^{48}$ &$1.43\times 10^{48}$ &$1.16\times 10^{49}$ &$5.39\times 10^{50}$ &$6.88\times 10^{50}$ &$1.54\times 10^{51}$ \\
\hline
0.58  &0.00 &0.00 &0.00 &$1.39\times 10^{47}$ &$6.38\times 10^{48}$ &$1.43\times 10^{48}$ &$1.16\times 10^{49}$ &$5.39\times 10^{50}$ &$6.88\times 10^{50}$ &$1.54\times 10^{51}$ \\
\hline
0.60  &0.00 &0.00 &$2.60\times 10^{47}$ &$1.39\times 10^{47}$ &$9.03\times 10^{49}$ &$1.43\times 10^{48}$ &$1.16\times 10^{49}$ &$5.39\times 10^{50}$ &$6.88\times 10^{50}$ &$1.54\times 10^{51}$ \\
\hline
0.62  &0.00 &0.00 &$2.60\times 10^{47}$ &$1.39\times 10^{47}$ &$9.03\times 10^{49}$ &$1.43\times 10^{48}$ &$1.16\times 10^{49}$ &$5.39\times 10^{50}$ &$6.88\times 10^{50}$ &$1.54\times 10^{51}$ \\
\hline
0.64  &0.00 &$2.22\times 10^{47}$ &$9.04\times 10^{47}$ &$1.60\times 10^{48}$ &$9.03\times 10^{49}$ &$1.43\times 10^{48}$ &$1.16\times 10^{49}$ &$5.39\times 10^{50}$ &$6.88\times 10^{50}$ &$1.54\times 10^{51}$ \\
\hline
0.66  &0.00 &$2.22\times 10^{47}$ &$9.04\times 10^{47}$ &$1.60\times 10^{48}$ &$9.03\times 10^{49}$ &$1.43\times 10^{48}$ &$1.16\times 10^{49}$ &$1.12\times 10^{51}$ &$6.96\times 10^{50}$ &$1.54\times 10^{51}$ \\
\hline
0.68  &0.00 &$2.22\times 10^{47}$ &$9.04\times 10^{47}$ &$1.60\times 10^{48}$ &$9.03\times 10^{49}$ &$1.43\times 10^{48}$ &$1.16\times 10^{49}$ &$1.12\times 10^{51}$ &$6.96\times 10^{50}$ &$1.54\times 10^{51}$ \\
\hline
0.70  &0.00 &$3.08\times 10^{48}$ &$1.57\times 10^{48}$ &$8.42\times 10^{48}$ &$9.03\times 10^{49}$ &$2.08\times 10^{50}$ &$1.16\times 10^{49}$ &$1.12\times 10^{51}$ &$6.96\times 10^{50}$ &$1.54\times 10^{51}$ \\
\hline
0.72  &0.00 &$3.08\times 10^{48}$ &$1.57\times 10^{48}$ &$8.42\times 10^{48}$ &$9.03\times 10^{49}$ &$2.08\times 10^{50}$ &$1.16\times 10^{49}$ &$1.12\times 10^{51}$ &$6.96\times 10^{50}$ &$1.54\times 10^{51}$ \\
\hline
0.74  &0.00 &$5.77\times 10^{48}$ &$5.15\times 10^{48}$ &$8.42\times 10^{48}$ &$9.03\times 10^{49}$ &$2.08\times 10^{50}$ &$5.88\times 10^{50}$ &$1.12\times 10^{51}$ &$6.96\times 10^{50}$ &$1.54\times 10^{51}$ \\
\hline
0.76  &0.00 &$5.77\times 10^{48}$ &$5.15\times 10^{48}$ &$1.54\times 10^{49}$ &$9.03\times 10^{49}$ &$2.08\times 10^{50}$ &$5.88\times 10^{50}$ &$1.12\times 10^{51}$ &$6.96\times 10^{50}$ &$1.54\times 10^{51}$ \\
\hline
0.78  &$3.72\times 10^{47}$ &$5.77\times 10^{48}$ &$5.15\times 10^{48}$ &$1.54\times 10^{49}$ &$9.03\times 10^{49}$ &$2.08\times 10^{50}$ &$5.88\times 10^{50}$ &$1.12\times 10^{51}$ &$6.96\times 10^{50}$ &$1.54\times 10^{51}$ \\
\hline
0.80  &$2.06\times 10^{48}$ &$1.53\times 10^{49}$ &$7.22\times 10^{48}$ &$1.54\times 10^{49}$ &$3.78\times 10^{50}$ &$2.08\times 10^{50}$ &$5.88\times 10^{50}$ &$1.12\times 10^{51}$ &$6.96\times 10^{50}$ &$1.54\times 10^{51}$ \\
\hline
0.82  &$2.06\times 10^{48}$ &$1.53\times 10^{49}$ &$7.22\times 10^{48}$ &$4.17\times 10^{49}$ &$3.78\times 10^{50}$ &$2.08\times 10^{50}$ &$5.88\times 10^{50}$ &$1.25\times 10^{51}$ &$6.96\times 10^{50}$ &$1.54\times 10^{51}$ \\
\hline
0.84  &$1.31\times 10^{49}$ &$4.39\times 10^{49}$ &$4.03\times 10^{49}$ &$4.17\times 10^{49}$ &$3.78\times 10^{50}$ &$3.87\times 10^{50}$ &$5.88\times 10^{50}$ &$1.25\times 10^{51}$ &$6.96\times 10^{50}$ &$1.54\times 10^{51}$ \\
\hline
0.86  &$1.48\times 10^{49}$ &$4.39\times 10^{49}$ &$4.03\times 10^{49}$ &$4.17\times 10^{49}$ &$3.78\times 10^{50}$ &$3.87\times 10^{50}$ &$5.88\times 10^{50}$ &$1.25\times 10^{51}$ &$6.96\times 10^{50}$ &$1.54\times 10^{51}$ \\
\hline
0.88  &$1.48\times 10^{49}$ &$4.39\times 10^{49}$ &$4.03\times 10^{49}$ &$1.78\times 10^{50}$ &$3.78\times 10^{50}$ &$3.87\times 10^{50}$ &$5.88\times 10^{50}$ &$1.25\times 10^{51}$ &$6.96\times 10^{50}$ &$1.54\times 10^{51}$ \\
\hline
0.90  &$1.75\times 10^{49}$ &$1.60\times 10^{50}$ &$1.37\times 10^{50}$ &$1.78\times 10^{50}$ &$3.78\times 10^{50}$ &$3.87\times 10^{50}$ &$5.88\times 10^{50}$ &$1.25\times 10^{51}$ &$6.96\times 10^{50}$ &$1.54\times 10^{51}$ \\
\hline
0.92  &$3.07\times 10^{49}$ &$1.60\times 10^{50}$ &$1.37\times 10^{50}$ &$1.78\times 10^{50}$ &$3.78\times 10^{50}$ &$3.87\times 10^{50}$ &$5.88\times 10^{50}$ &$1.25\times 10^{51}$ &$6.96\times 10^{50}$ &$1.54\times 10^{51}$ \\
\hline
0.94  &$3.07\times 10^{49}$ &$2.62\times 10^{50}$ &$1.81\times 10^{50}$ &$2.68\times 10^{50}$ &$3.78\times 10^{50}$ &$3.87\times 10^{50}$ &$5.88\times 10^{50}$ &$1.25\times 10^{51}$ &$6.96\times 10^{50}$ &$1.54\times 10^{51}$ \\
\hline
0.96  &$7.87\times 10^{49}$ &$2.62\times 10^{50}$ &$1.81\times 10^{50}$ &$2.68\times 10^{50}$ &$3.78\times 10^{50}$ &$3.87\times 10^{50}$ &$5.88\times 10^{50}$ &$1.25\times 10^{51}$ &$6.96\times 10^{50}$ &$1.54\times 10^{51}$ \\
\hline
0.98  &$7.87\times 10^{49}$ &$2.62\times 10^{50}$ &$1.81\times 10^{50}$ &$2.68\times 10^{50}$ &$3.78\times 10^{50}$ &$3.87\times 10^{50}$ &$5.88\times 10^{50}$ &$1.25\times 10^{51}$ &$6.96\times 10^{50}$ &$1.54\times 10^{51}$ \\
\hline
\hline
\end{tabular}
\end{table*}

\section{Numerical Methodology}
\subsection{Basic Model}\label{method}
All simulations presented in this paper are carried out with the publicly available Enzo code \citep{Enzo}. Enzo solves the Eulerian equations of cosmological radiation hydrodynamics using the Particle-Mesh method for the dark matter with CIC interpolation \citep{HockneyEastwood88},  a dual-energy formulation of the Piecewise Parabolic Method for the gas \citep{Bryan95}, and FFTs for the gravitational field. We use it in its 6-species primordial gas chemistry mode, wherein the nonequilibrium evolution of H, H$^+$, He, He$^+$, He$^{++}$ and e$^-$ is computed using the backward difference formula (BDF) solver of \cite{Anninos97}.  We use Enzo's built-in implicit flux-limited diffusion (FLD) radiative transfer solver \citep{ReynoldsEtAl2009,Norman13} for the transfer of ionizing photons, which are treated in the grey approximation. In \cite{Norman13} we show that Enzo's FLD and Moray ray tacing radiative transfer solver \citep{Wise11_Moray} give nearly identical ionization histories in a reionization test problem.  We use FLD because it is much faster, especially as the volume becomes fully ionized. The ionizing sources are assumed to be low metallicity star forming galaxies in halos of mass $\mh \geq 10^7\Ms$. Thus we ignore the radiative contribution of Pop III stars. We return to the impact of this assumption in the discussion section. The spectral energy distribution (SED) of the stellar radiation is the same as in \cite{So14}, which is the SED derived by \cite{Ricotti02} for a $Z=0.04 Z_{\odot}$ stellar population but truncated above 4 Ryd.  We also include in the simulation star formation and supernova feedback using the simple parameterized model of \cite{So14}, however we do not use it for calculating radiative feedback because of the improved, halo-based, resolution-insensitive model introduced here. Because we input into the simulation the number of escaping ionizing photons measured at the virial radius from the {\it Renaissance Simulations} (Sec. 2), we do not need to assume an ionizing escape fraction. Therefore $f_{esc}$ is not a parameter of the simulation, as it is in previous reionization simulations which relate halo emissivities to halo mass in a parameterized way. Here we use the results of higher resolution simulations which provide these ionizing emissivities directly. Our approach is generally applicable. We also only need to resolve the virial radii of the halos of importance, which greatly relaxes the spatial resolution requirement, but not the mass resolution (see below).  A WMAP7 $\Lambda$CDM cosmological model is used: $\Omega_M=0.27$, $\Omega_\Lambda=0.73$, $\Omega_b=0.047$, $h=0.7$, $\sigma_8=0.82$, and $n=0.95$, where the parameters have the usual
definitions. All simulations start from redshift 99 and run until the simulation volume is fully ionized. 

In this paper we present three simulations differing only in mass and spatial resolution and box size. All use inline halo finding and assign emissivities to halos as described in Secs. \ref{halofinding} and \ref{input}. Their properties are summarized in Table \ref{tab_sim}. The first two constitute a resolution study, which show the importance of including halos as small as $10^7\Ms$. 256\_all is a $256^3$ cell/particle simulation in a 6.4 comoving Mpc box. This is the same as the test problem presented in \cite{So14} and \cite{Norman13}, and has the same mass and spatial resolution as the science run analyzed in \cite{So14}. In that case the dark matter particle mass was chosen so that the halo mass function was complete above $\mh = 10^8 \Ms$. 512\_all is a $512^3$ cell/particle simulation in the same box. It has 2 times the spatial resolution and 8 times the mass resolution as the 256\_all simulation. The halo mass function is complete to $ 10^{6.8}M_{\odot}$, defined as halos containing at least 100 particles, essential for including the MCs. These simulations are discussed in Sec. \ref{resolution}. The third, 1152\_all simulation, is our science run. It is a $1152^3$ cell/particle simulation in a 14.4 comoving Mpc box. It has the same mass and spatial resolution as 512\_all simulation except in a box $2.25$ as large ($11.4$ times the volume). This simulation is discussed in Sec. \ref{science_run}. 

\subsection{Inline Halo Finding}\label{halofinding}
The Enzo code has an embedded Python capability which allows the user to execute Python scripts which operate on Enzo's internal data structures as the code runs. As the {\tt yt} toolkit \citep{yt_full_paper} is implemented in Python, many of {\tt yt}'s analysis capabilities can be run inline with the computation. This includes the Parallel HOP halo finder \citep{Skory10}, which we employ here. 
In all simulations in Table \ref{tab_sim}, we use embedded Python to call a script every 20 million years to do the following: (i) find halos and calculate their integral properties, including their virial masses;  (ii) assign an emissivity to each halo according to its mass (Section \ref{input}), (iii) zero the old emissivity field array, and use the halos' positions and emissivities to compute a new emissivity field array (Section \ref{input}). We choose 20 million years since that is the typical lifetime of OB stars in a coeval stellar population. The emissivity fields are then kept constant until the next inline Python script is called 20 million years later. After clearing all the old emissivity fields, we distribute the new halo emissivity evenly into 27 adjacent cells (a 3$\times$3$\times$3 cube) centered at the cell that contains the halo's center of mass. Since $\dot{N}_{\rm ion,esc}$ (Equation \ref{eqn:emis9} and Table \ref{tab_sim}) already include escaping fractions, by distributing the emissivity field in a larger region instead of the center cell we avoid the ionizing photons being absorbed again. 

In \cite{Xu16}, escaped ionizing photons are measured at the halo's virial radius, which for the masses and redshifts we are concerned with, ranges from 0.3 to 1.7 proper kpc/h. The grid resolution used here is 12.5 comoving kpc, which is about 1 proper kpc at $z \sim 10$. By injecting photons into a 3x3x3 cube centered on the halo, we are in effect injecting ionizing photons at about 2-3 times $r_{vir}$. We are assuming no additional absorption in the small range of unresolved intermediate scales, an assumption borne out by tests carried out by \cite{Wise14} whose method we employ. 

Because low mass halos ($10^7M_\odot<\mh<10^{8.5}M_\odot$) do not have a unit probability of actively forming stars (\citet{Xu16}, Table \ref{tab_hao}), halos emitting during this 20 million years will likely not emit for the next 20 million years. This is consistent with the results of \cite{KimmCen14}, who find bursty star formation with a duty cycle of about 20 Myr in their simulations of EoR galaxies. In this way we take the intermittency of the contribution from low mass halos into consideration. 

\begin{table*}
\centering
\caption{Summary of simulations}
\label{tab_sim}
\begin{tabular}{cccccccc}
\hline\hline
Simulation & L$_{box}$ (Mpc) & $N_{cell}=N_{p}$ & $m_p (M_{\odot})$ & $100m_p (M_{\odot})$  & $\Delta x (kpc)$ &  MC halos included & $z_{ov}$  \\
\hline
256\_all & 6.4 & 256$^3$ & $4.8 \times 10^5$  & $4.8 \times 10^7$ & 25 & N & 5.56  \\
\hline
512\_all & 6.4 & 512$^3$ & $6 \times 10^4$ & $6 \times 10^6$ & 12.5 & Y & 5.80  \\
\hline
1152\_all & 14.4 & 1152$^3$ & $6 \times 10^4$ & $6 \times 10^6$ & 12.5 & Y & 7.08  \\
\hline
\hline
\end{tabular}
\parbox[t]{0.73\textwidth}{\textbf{Notes.} 
Box size and resolution element are comoving. $z_{ov}$ is overlap redshift; i.e., when 100\% of the volume is ionized to above 99.9\% ionization fraction. 
}
\end{table*}

\subsection{Assigning the Emissivity Field}\label{input}
Enzo's flux-limited diffusion radiation transport solver is sourced by an emissivity field $\eta(\vec{x})$ defined on the mesh, where [$\eta$] = erg/s/cm$^3$ \citep{Norman13}. To generate this field, we loop over the entire mesh, and sum the halo ionizing photon rates in each cell. As previously described, the photons are then equally distributed to a 3$\times$3$\times$3 block of cells centered on each emitting cell.  To convert to an energy emissivity, we multiply the photon flux by the group average photon energy of 21.6 eV and divide by the cell volume.
Instead of using a constant mass-to-light ratio \citep{Iliev07}, we assign emissivities to halos according to the $\dot{N}_{\rm ion,esc}-\mh$ relations derived from the Renaissance Simulations \citep{Xu16}, as previously discussed. In all simulations, for the larger HMACH halos ($\mh>10^9M_\odot$), we use Equation \ref{eqn:emis9} to get their $\dot{N}_{\rm ion,esc}$. For each halo with mass between $10^7$ to $10^9$ $M_\odot$, we generate a random number between 0 and 1 and use it to choose the corresponding $\dot{N}_{\rm ion,esc}$ from Table \ref{tab_hao} in its mass bin. Note that the majority of the cells in Table 1 with mass below $10^{8.4}M_\odot$ have zero emissivity. For example, for halos with mass around $10^7M_\odot$, there are only less than 2\% of them with nonzero emissivity. This is due to the inefficiency of star formation and the supernova feedback in low mass halos \citep{Wyithe13}. Each halo larger than $10^{8.6}M_\odot$ has nonzero emissivity, but the value may change every 20 million years when a new value is chosen in its mass bin or when it falls into another mass bin. 

\section{Results}

\subsection{Resolution study -- 256$^3$ and 512$^3$ simulations}\label{resolution}
To understand the role of the smallest galaxies in reionization we performed two simulations, 256\_all and 512\_all, both with 6.4 comoving Mpc per side. As described above, the 512\_all simulation has twice the spatial resolution and eight times the mass resolution as the 256\_all simulation. 
These two simulations thus constitute a small resolution study. The box size and resolution were chosen so that the halo mass functions are complete to $10^8\Ms$ and $10^7\Ms$, respectively. Here completeness means halos of these masses have at least 100 dark matter particles. 

\begin{figure}
\includegraphics[width=1.05\columnwidth]{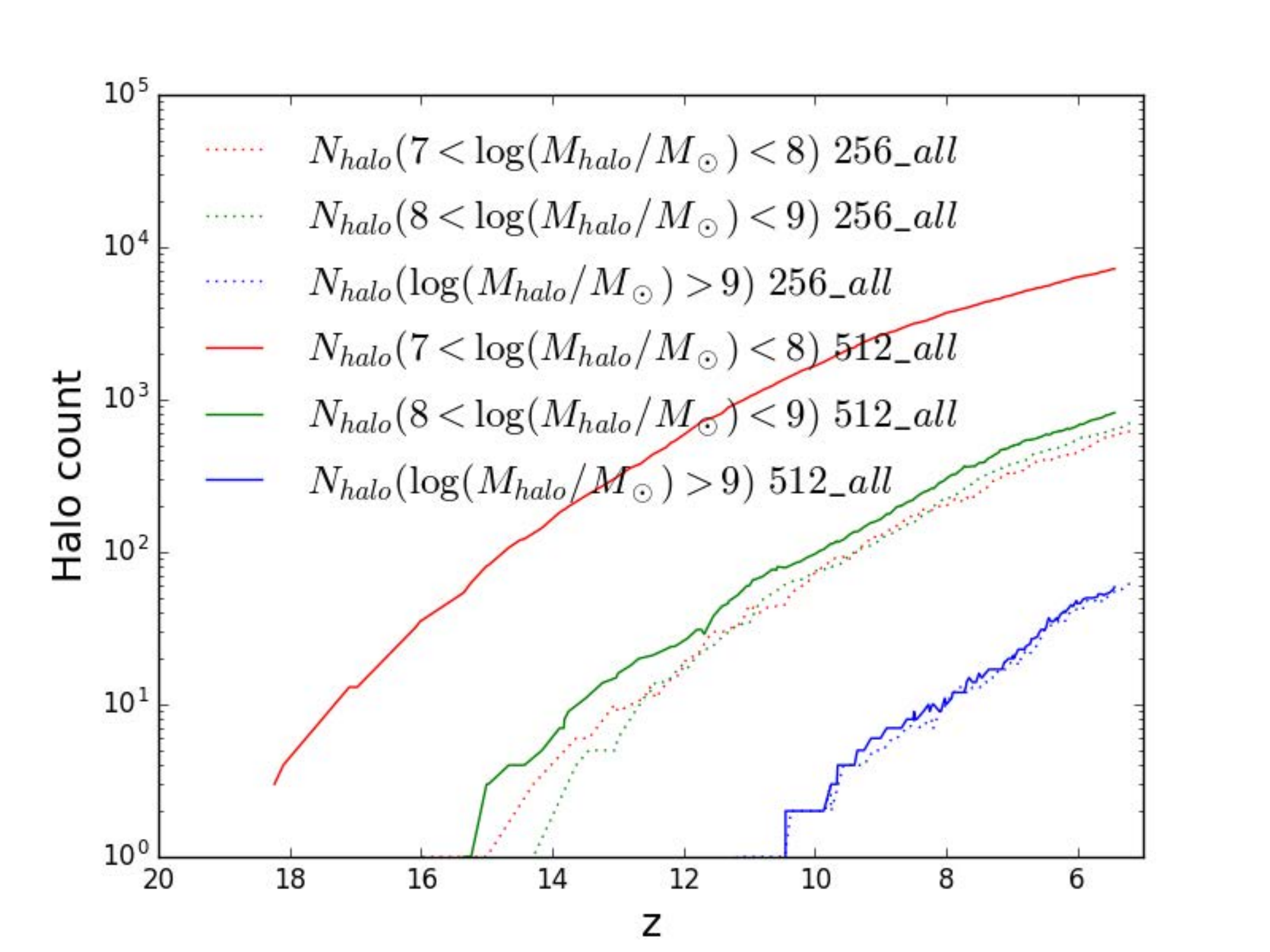}
\caption{Halo counts vs. redshift in the entire volume for the 256\_all and 512\_all resolution study simulations for three mass bins: $10^7 \leq \mh/M_{\odot} \leq  10^8$; $10^8 \leq \mh/M_{\odot}  \leq 10^9$; and $\mh/M_{\odot} > 10^9$. Note the virtual absence of halos in the lowest mass bin corresponding to metal-line cooling halos (MCs) in the lower resolution simulation 256\_all. Warren fits to the halo counts in mass bins are plotted as solid lines.  }
\label{256_512_halos}
\end{figure}

Figure \ref{256_512_halos} shows the halo counts versus redshift in 3 mass bins in both simulations: MCs - $10^7 \leq \mh/M_{\odot} \leq  10^8$; LMACHs - $10^8 \leq \mh/M_{\odot}  \leq 10^9$; and HMACHs - $\mh/M_{\odot} > 10^9$. 
Referring to the 512\_all curves, we see MCs, LMACHs, and HMACHs begin forming at $z \sim18, 15, 11$, respectively. Comparing these curves to their 256\_all counterparts, we see that the LMACH and HMACH formation histories are converged, but the MCs are severely underestimated. 
Due to the higher mass resolution, the MC halo counts in 512\_all are 1.5 to 2 orders more than those in 256\_all, and they begin forming sooner. The first halo with $\mh>10^7M_\odot$ appears at redshift $\sim$18.2 in 512\_all, which is earlier than redshift $\sim$16.0 in 256\_all. 

\begin{figure}
\includegraphics[width=0.95\columnwidth]{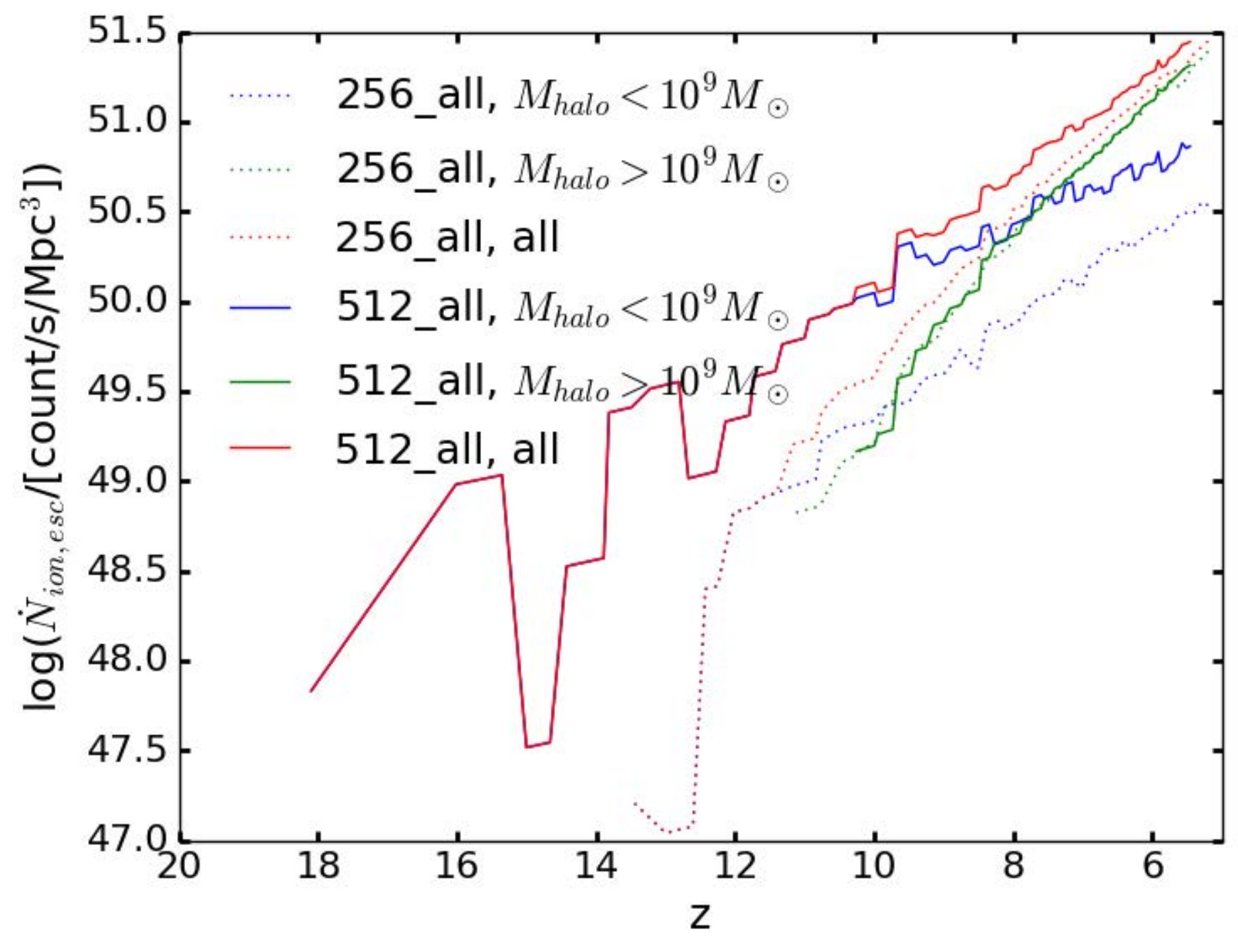}
\includegraphics[width=0.95\columnwidth]{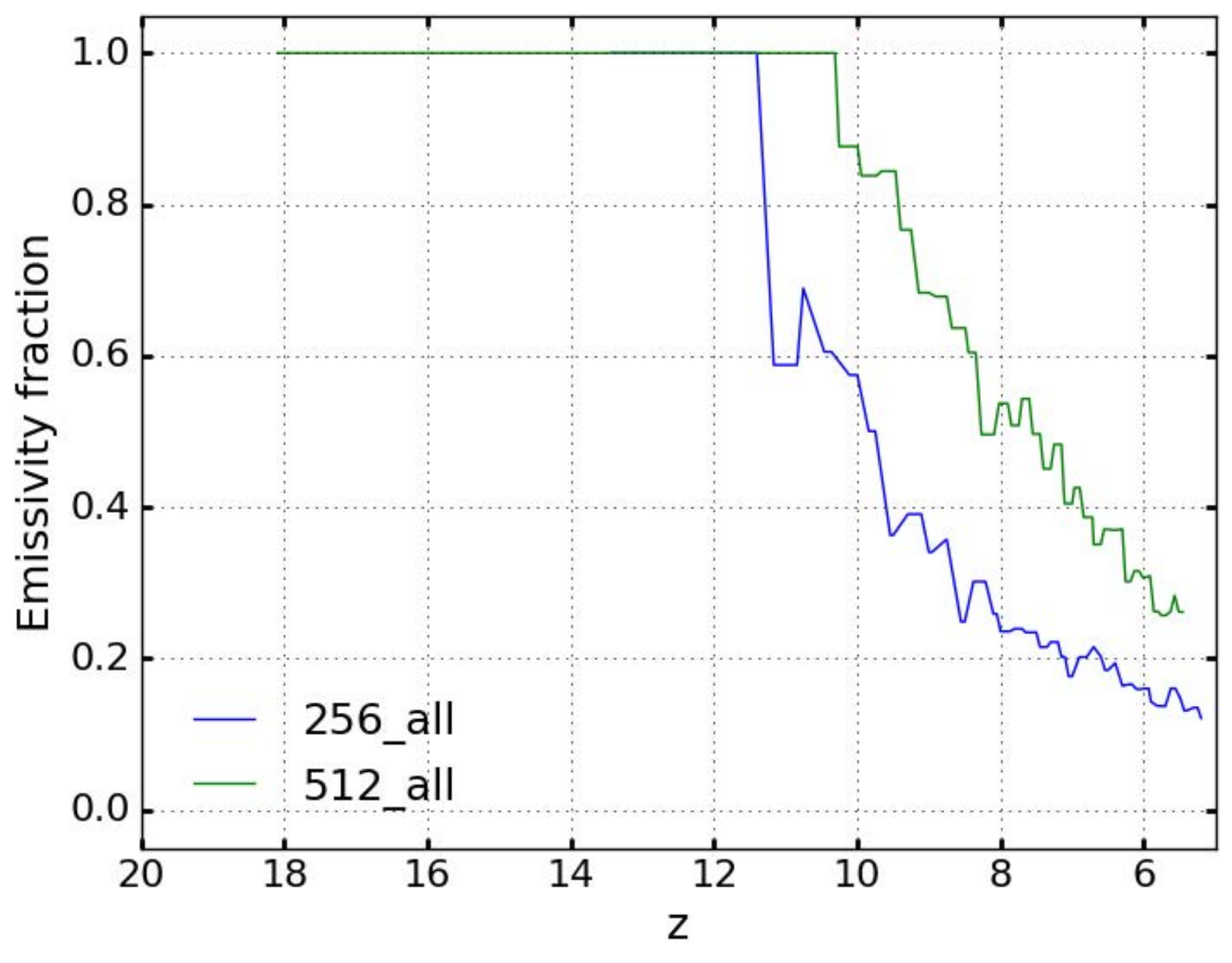}
\caption{Contribution of halos of different masses to the ionizing emissivity as a function of redshift. Top: comoving ionizing luminosity density vs. redshift from halos below and above $10^9\Ms$ and their sum, for the 256\_all test simulation (solid lines) and 512\_all test simulation (dotted lines). Bottom:  Fraction of ionizing photons from halos below $10^9\Ms$ versus redshift for 256\_all and 512\_all. The fraction drops as HMACHs become the dominant ionizing sources.}
\label{256_512_emis_frac}
\end{figure}

Figure \ref{256_512_emis_frac} shows the evolution of the ionizing photons emitted per comoving cubic Mpc as a function of redshift in the 256\_all and 512\_all simulations. In the top panel we show the comoving ionizing luminosity density from halos below and above $\mh = 10^9\Ms$, as well as the total; in the bottom panel we show the fraction of the total ionizing luminosity coming from halos below $\mh = 10^9\Ms$.  Looking at the top panel we can see that the HMACH contribution becomes dominant below $z \sim 10$ in the 256\_all simulation, but not until $z \sim 8$ in the 512\_all simulation . Because the HMACH population is virtually identical in both simulations, the difference is due to the enhanced contribution of the low mass halos, and specifically the MCs since the LMACH populations are also virtually identical in the two simulations (Fig. \ref{256_512_halos}).  Referring to the bottom panel, when there are no HMACHs, all the ionizing photons come from low mass halos so the ratio is one. The ratio drops below unity when the first HMACH halo forms, which occurs at slightly different redshifts due to a resolution effect. At $z \approx 11.5$, when the first HMACH halo forms in the lower resolution simulation, it is resolved into two lower mass halos about to merge in the higher resolution simulation. Then as more HMACHs form and become dominant the ratio drops to $\sim$15\%(25\%) when the reionization completes in 256\_all(512\_all), with some fluctuations in between. The ratio is always higher in 512\_all because it has more low mass halos than 256\_all but about the same amount of HMACHs.  

\begin{figure}
\includegraphics[width=\columnwidth]{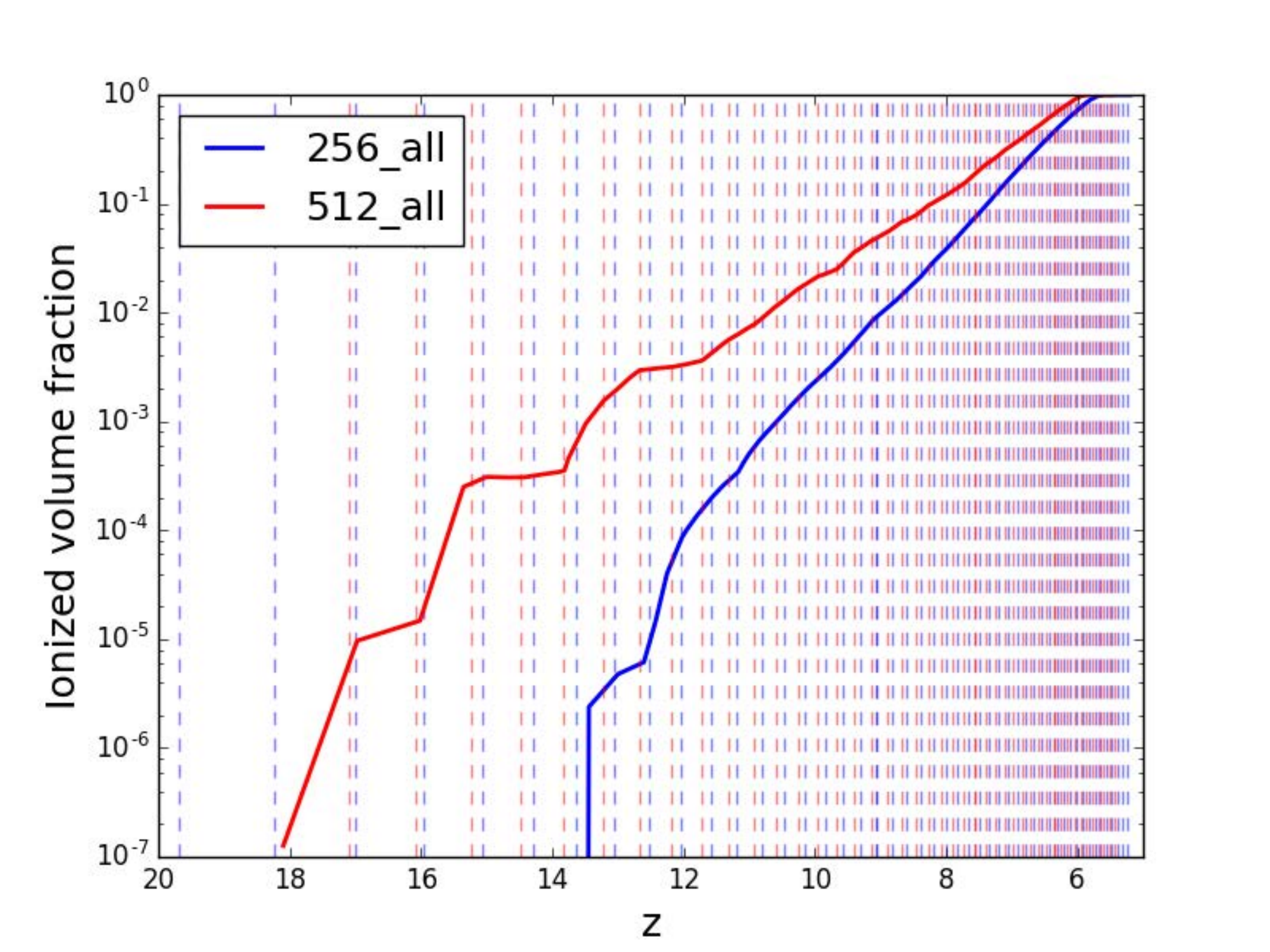}
\caption{The ionized volume fraction as a function of redshift for 256\_all and 512\_all. The vertical dashed lines are approximately 20 million years apart indicating the time when inline Python works to find halos and assign ionizing emissivities according to the current halos' mass as described in Sec. \ref{input}.}
\label{256_512_ion_frac10}
\end{figure}

Figure \ref{256_512_ion_frac10} shows the evolution of the volume fraction ionized above an ionization fraction of 10\% for the 256\_all and 512\_all simulations. The vertical dashed lines indicate when the emissivity field is reset every 20 Myr. Although the first halo with $\mh>10^7M_\odot$ appears at redshift $\sim$16.0 in 256\_all,
the volume doesn't begin to ionize until redshift $\sim$13.5. This is due to the low probability for lower mass halos to emit (Table \ref{tab_hao}). In 512\_all the time between when the first low mass halo emits and when the first HMACH emits is longer than in 256\_all, and there are several "stair steps". This is also due to the randomness in the turning on and off of low mass halos. When more halos are turning off, there would be a relatively flat part in the ionized fraction curve. Interestingly, the 512\_all simulation completes reionization slightly sooner than the 256\_all simulation, this despite the fact that HMACHs have dominated the photon budget by then. This result can be understood as a simple consequence that reionization completion depends on the total number of ionizing photons, which is higher for all redshifts in the 512\_all simulation as compared to the 256\_all simulation (Fig. \ref{256_512_emis_frac}).

\begin{figure*}
\begin{center}
\centerline{
\mbox{\includegraphics[width=0.95\columnwidth]{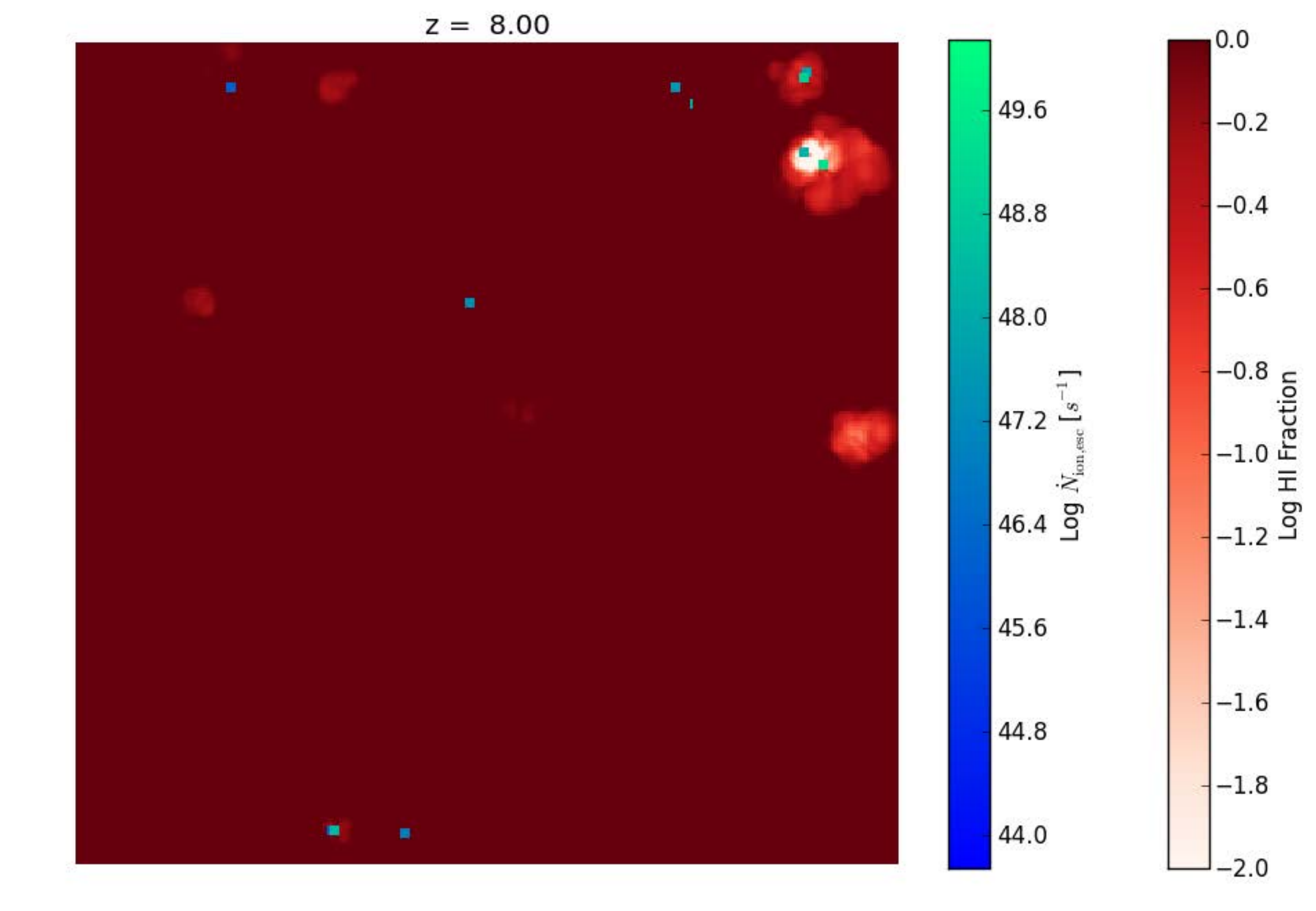}}
\mbox{\includegraphics[width=0.95\columnwidth]{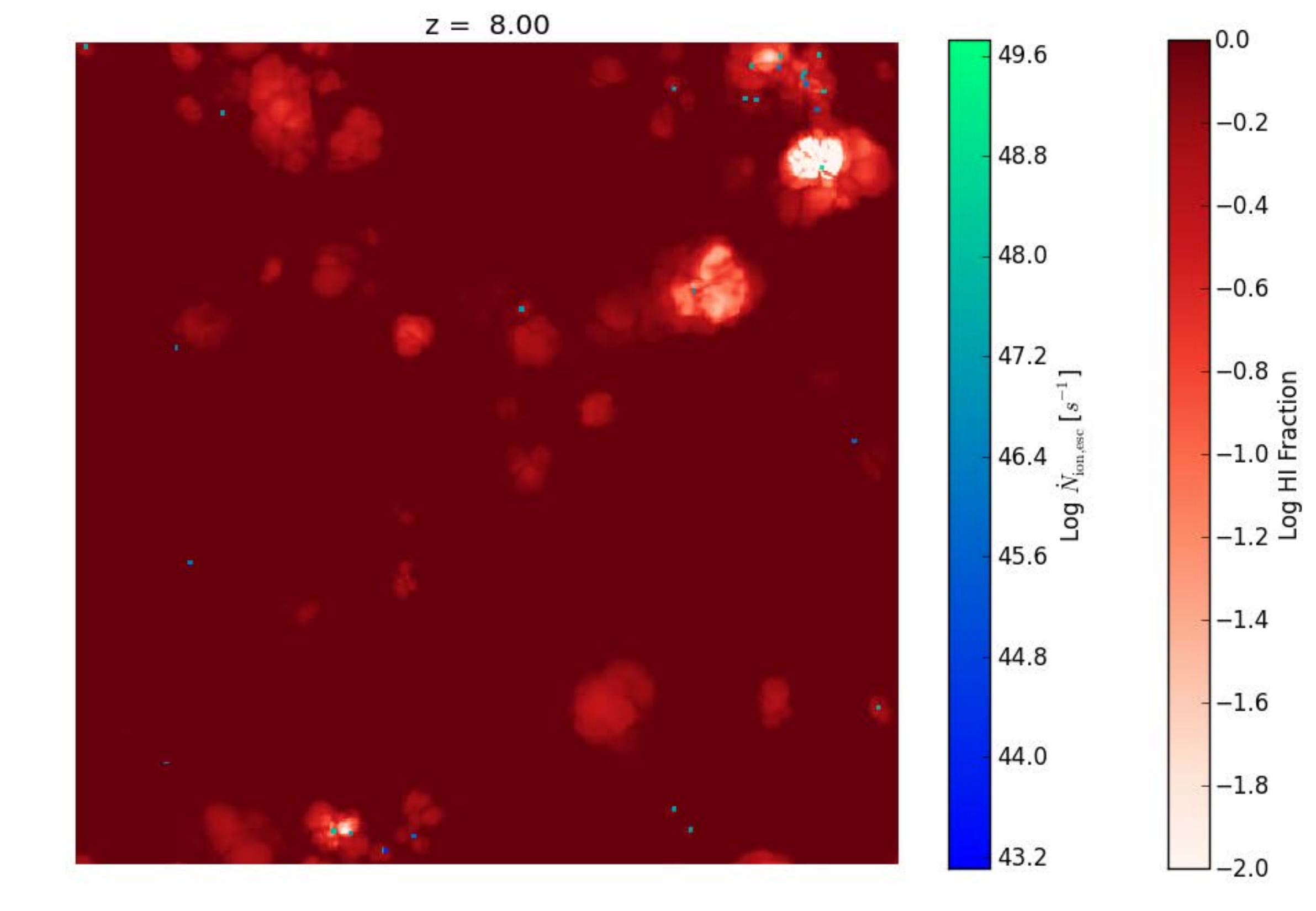}}}
\centerline{
\mbox{\includegraphics[width=0.95\columnwidth]{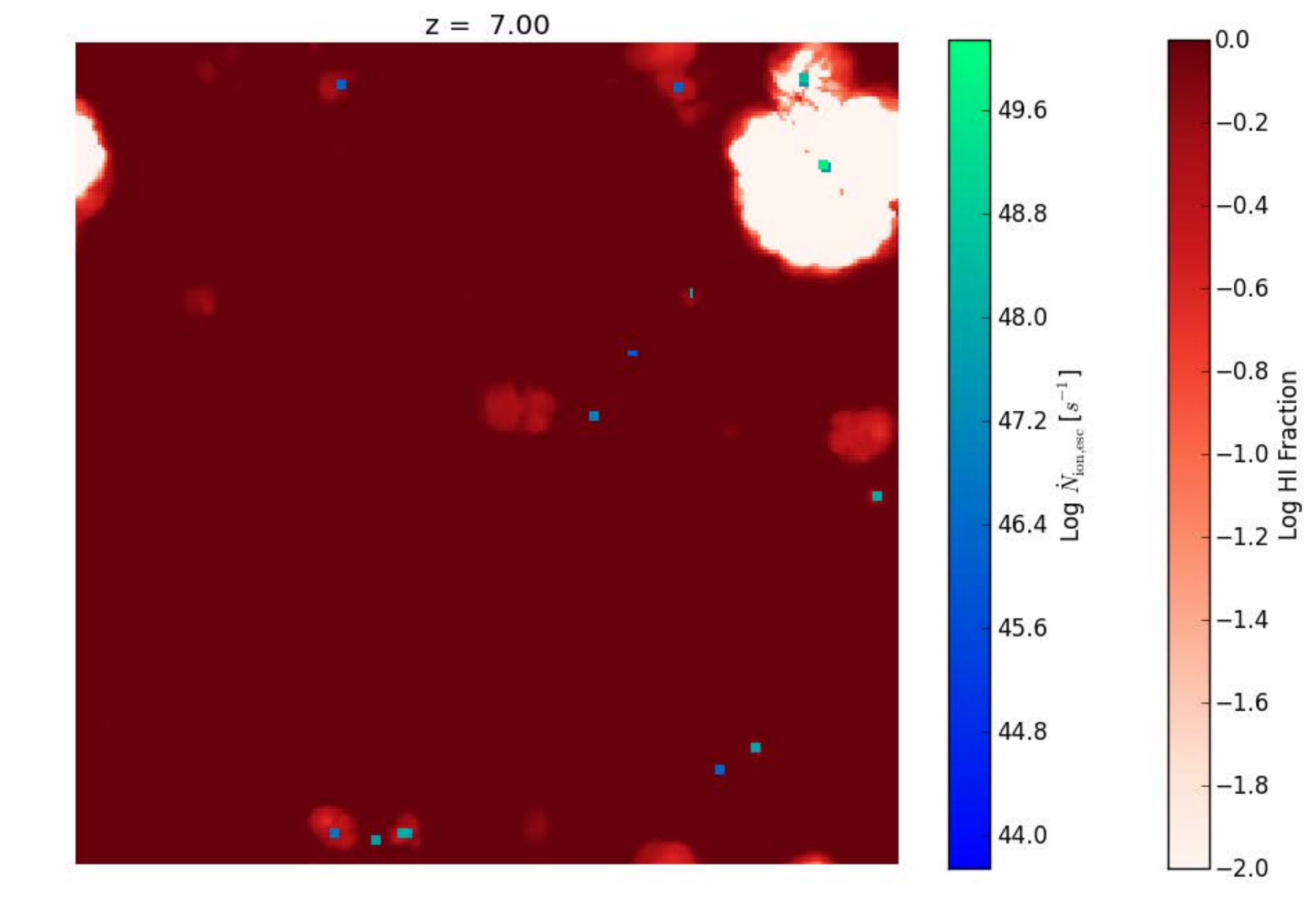}}
\mbox{\includegraphics[width=0.95\columnwidth]{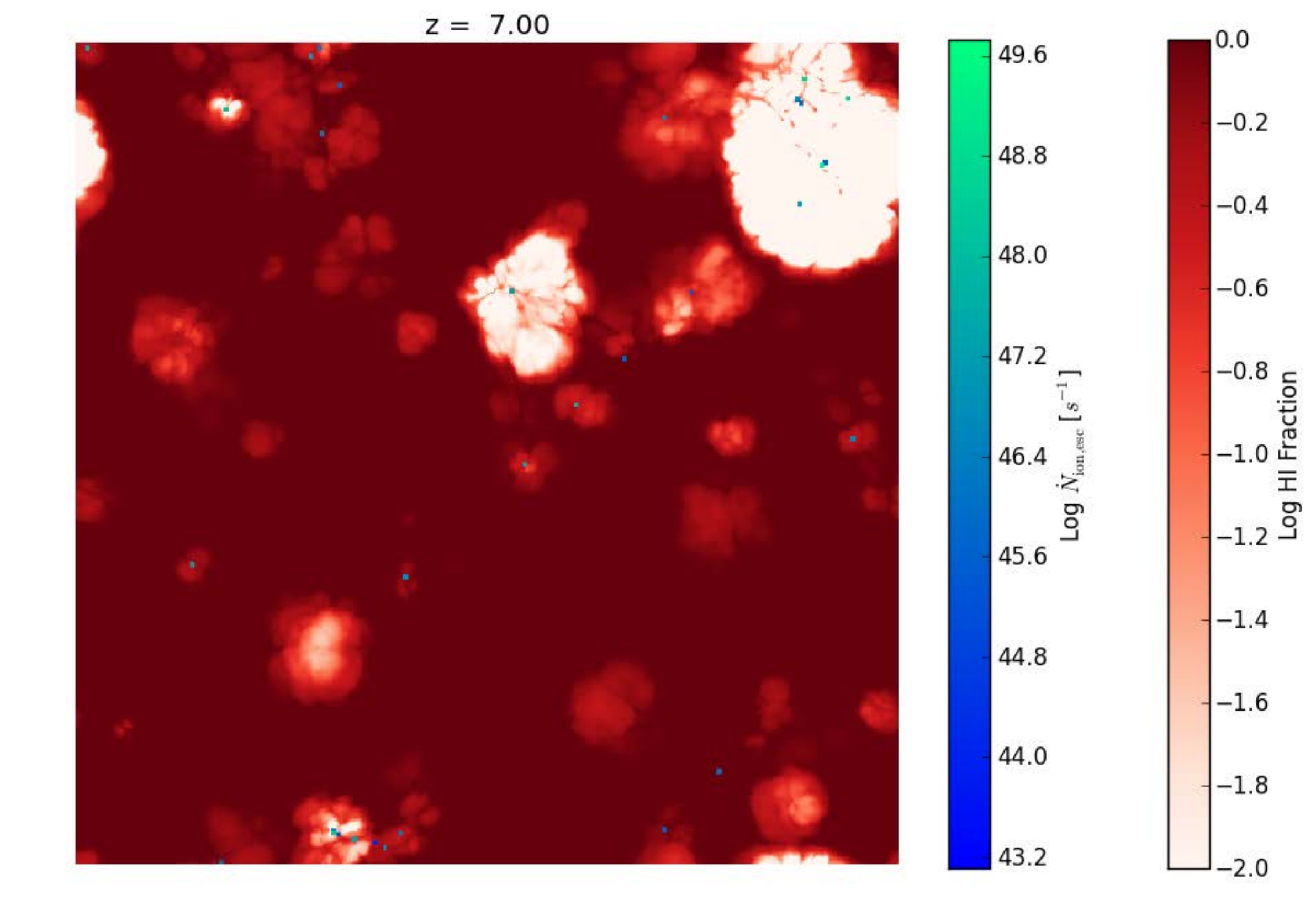}}}
\centerline{
\mbox{\includegraphics[width=0.95\columnwidth]{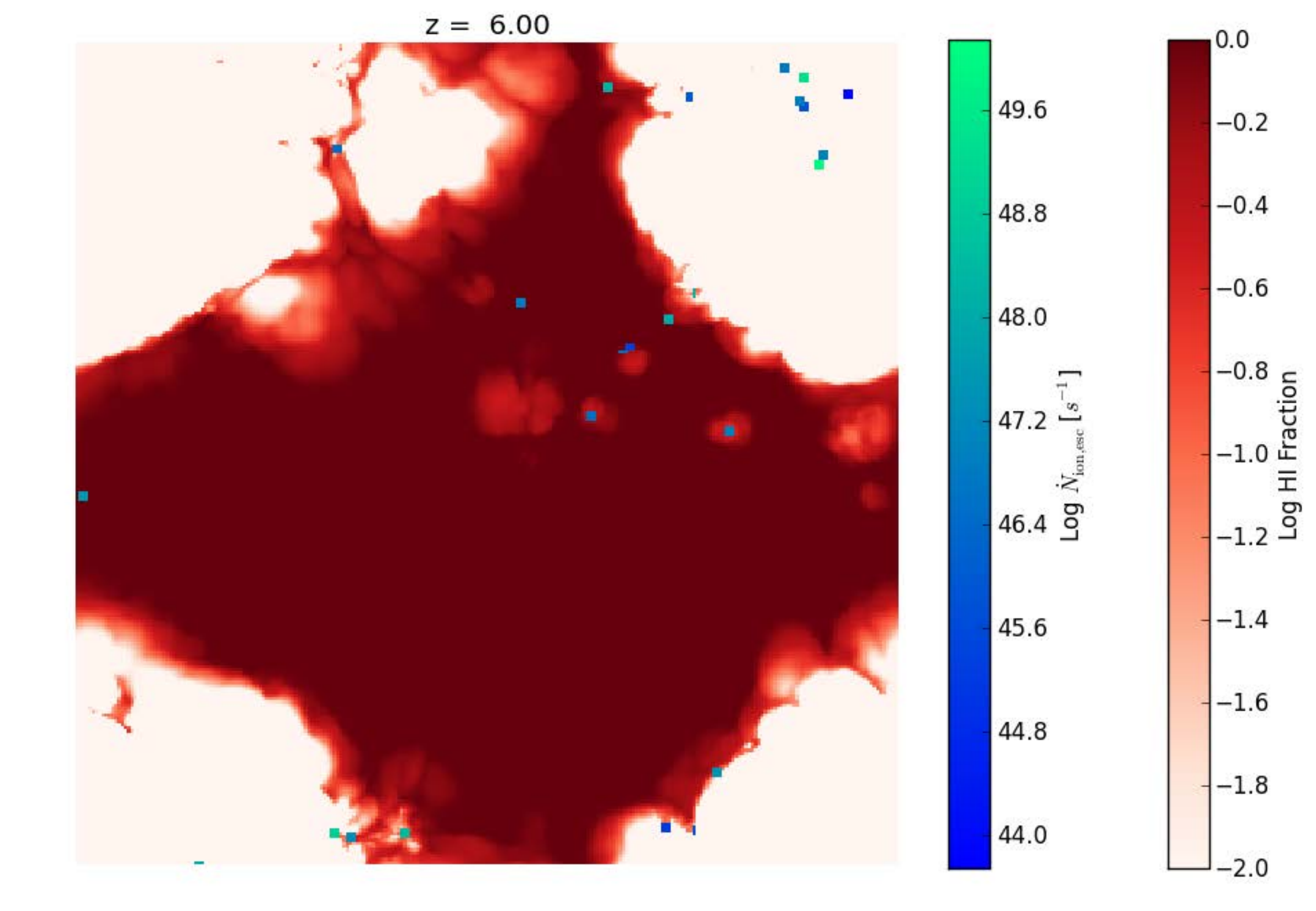}}
\mbox{\includegraphics[width=0.95\columnwidth]{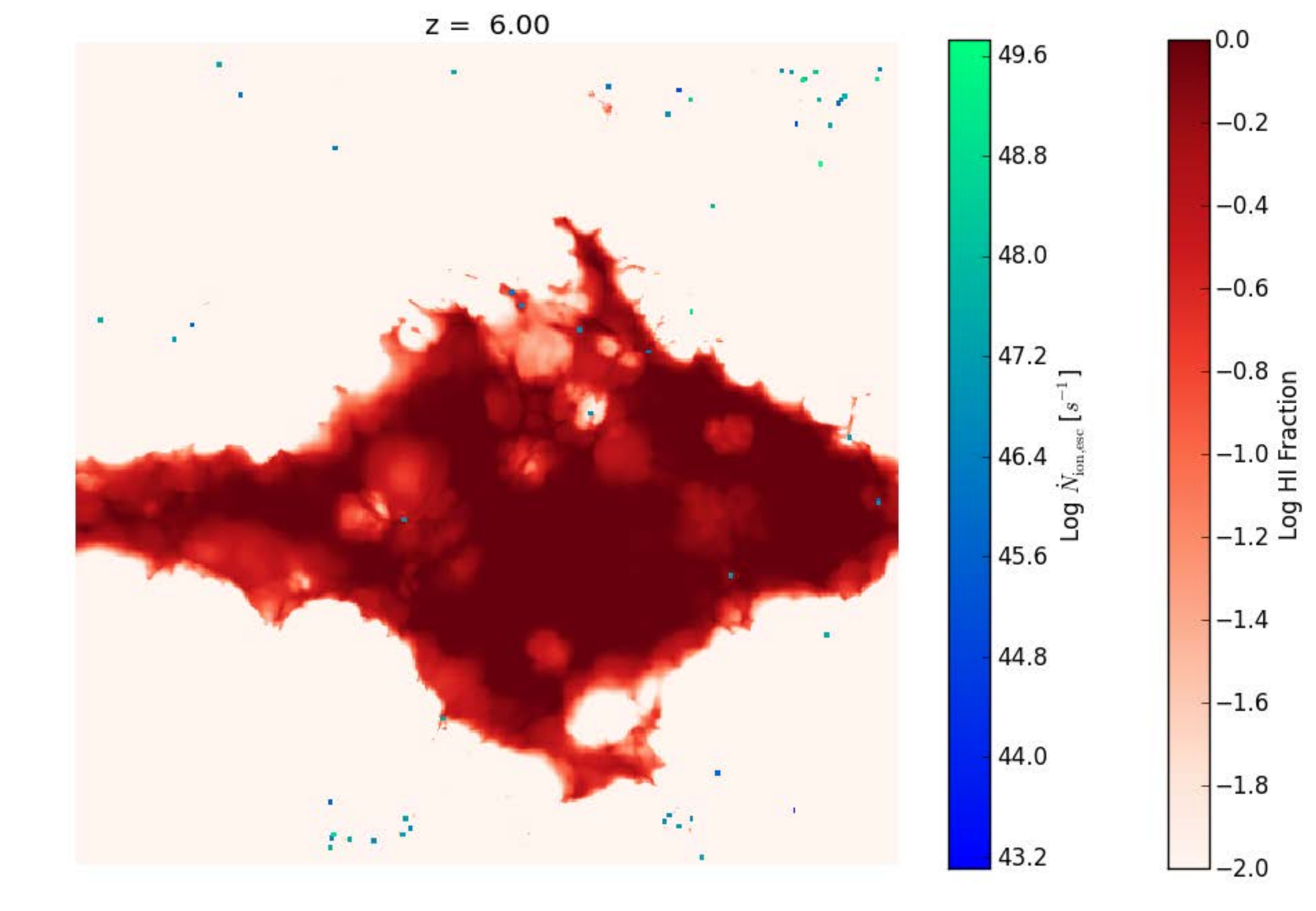}}}
\end{center}
\caption[Projections of the logarithm of the neutral hydrogen fraction for the $256^3$ and $512^3$ test simulations] {Volume-weighted projections of the logarithm of the neutral hydrogen fraction through the 6.4 Mpc box at redshifts $z=8, 7$ and $6$ for the $256^3$ (left column) and $512^3$ (right column) test simulations. Note the increase in the number of relic H II regions in the high resolution simulation. The projected ionizing emissivity field $\eta$ which sources the radiation transport solver is superimposed. Its colorbar has been converted to $\dot{N}_{ion,esc}=\eta V_{cell}/\bar{e}_{\gamma}$, where $V_{cell}$ is the cell volume, and $\bar{e}_{\gamma}$ is the mean energy per photon, 21.6 eV. }
\label{256_512_comparison}
\end{figure*}

To complete our presentation of the resolution study results, we show in Fig. \ref{256_512_comparison} side-by-side volume-weighted projections of the logarithm of the neutral hydrogen fraction through the 6.4 Mpc volume at $z=8, 7$ and $6$. The included color bar is chosen to show highly ionized gas as white, and partially ionized gas as shades of red-brown. The superimposed colored pixels is a projection of the instantaneous ionizing emissivity field $\eta$ computed by binning emitting halos on the Eulerian mesh (Sec. 3.3). For ease of comprehension, in the color bar we have converted $\eta$, which has units of erg/s/cm$^3$ to $\dot{N}$, the ionizing photon flux produced by halos in that cell by multiplying $\eta$ by the cell volume and dividing by the mean energy per photon 21.6 eV. One can see the larger number of smaller \hii~ regions at earlier redshifts in the higher resolution simulation, as compared to the lower resolution simulation. Many of these are relic \hii~ regions as their sources have turned off according to our probabilistic model of star formation in low mass halos. One can also see that reionization has progressed further by $z=6$ in the high resolution simulation, and that the strong ionization front driven by sustained star formation in the upper right corner of the cube is sweeping over smaller active and relic \hii~ regions from earlier star formation in smaller halos. We explore this topic more thoroughly in the next section.

\subsection{Science run--larger high-resolution volume}\label{science_run}

\begin{figure*}
\begin{center}
\centerline{
\mbox{\includegraphics[width=0.95\columnwidth]{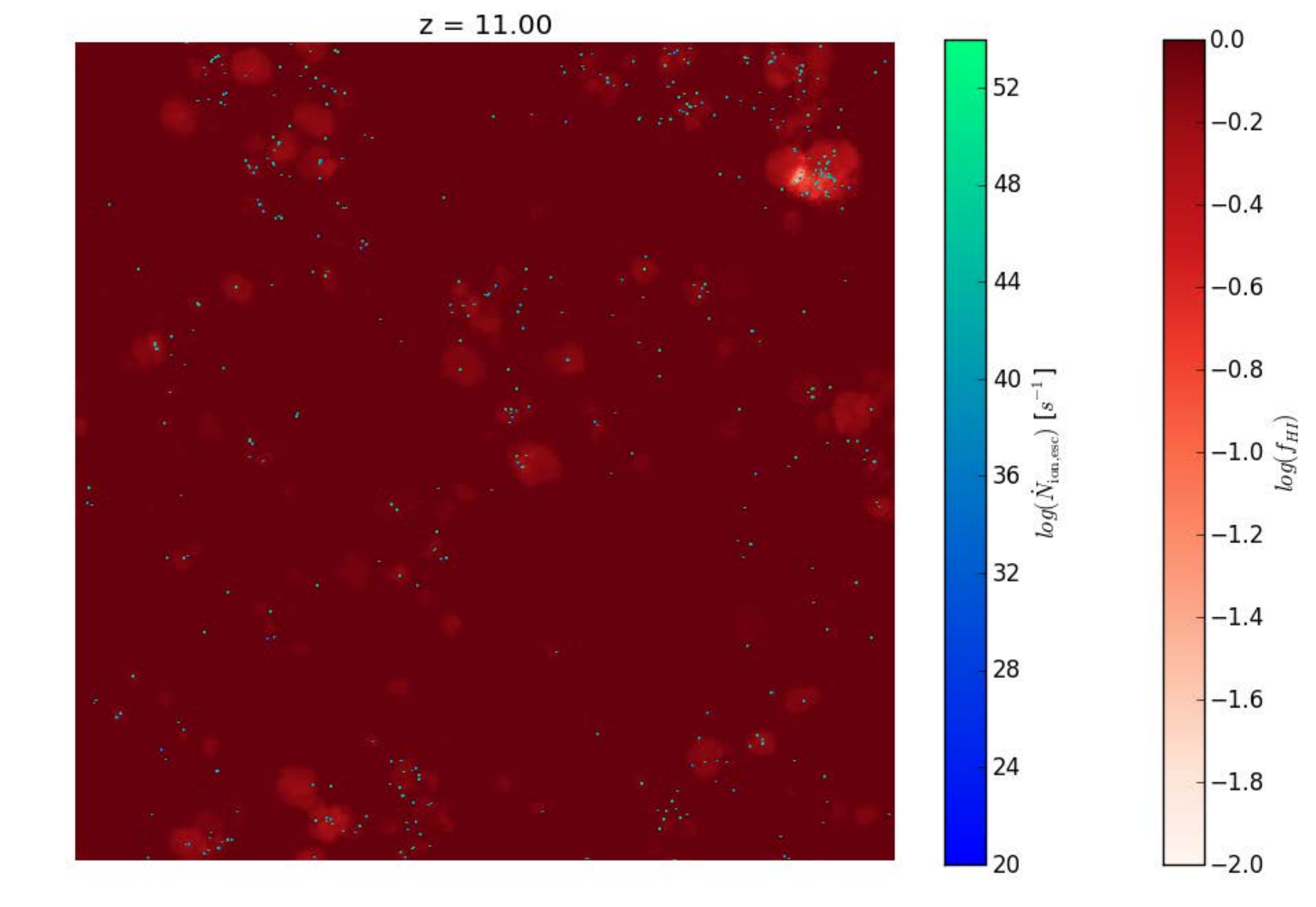}}
\mbox{\includegraphics[width=0.95\columnwidth]{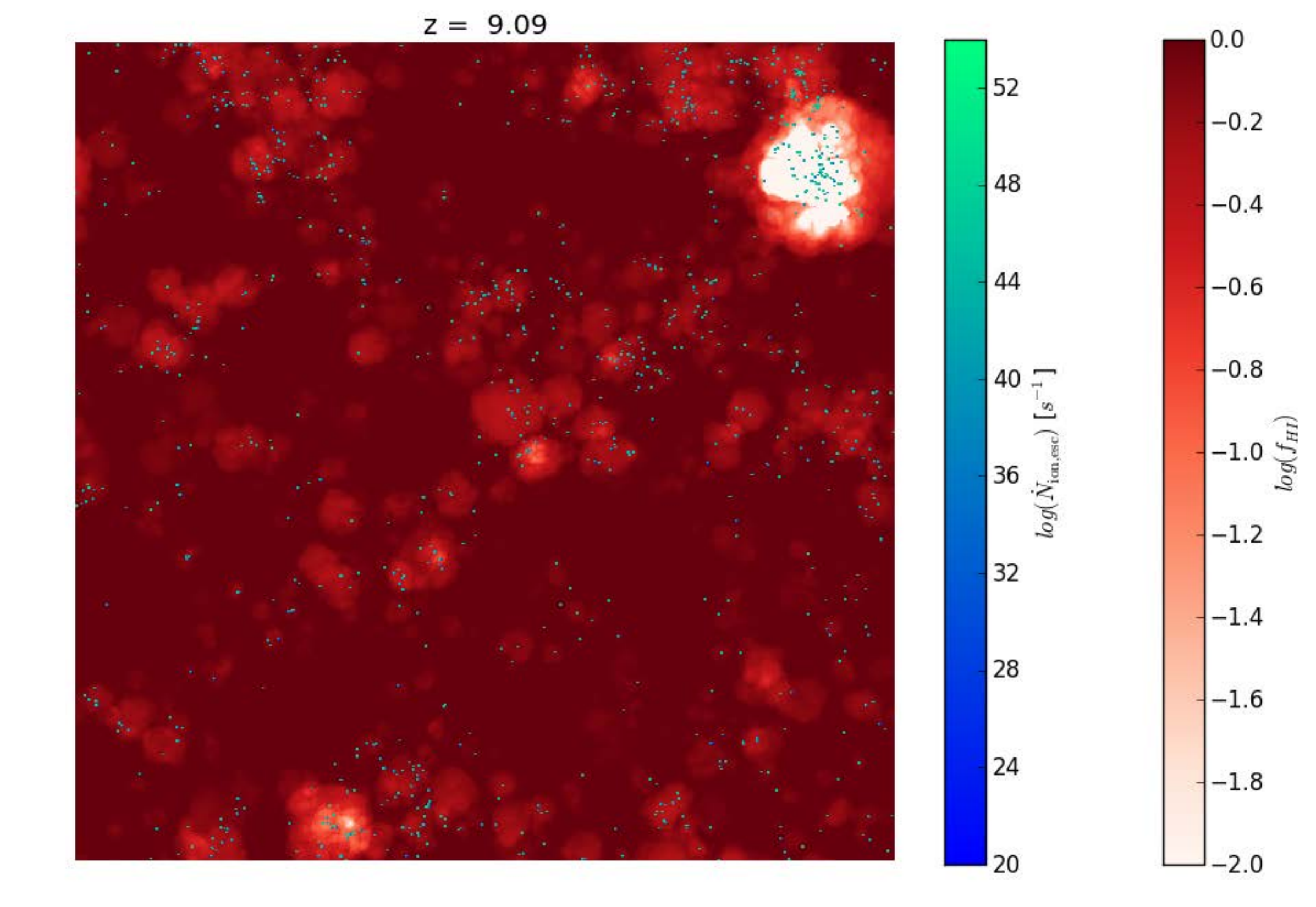}}}
\centerline{
\mbox{\includegraphics[width=0.95\columnwidth]{{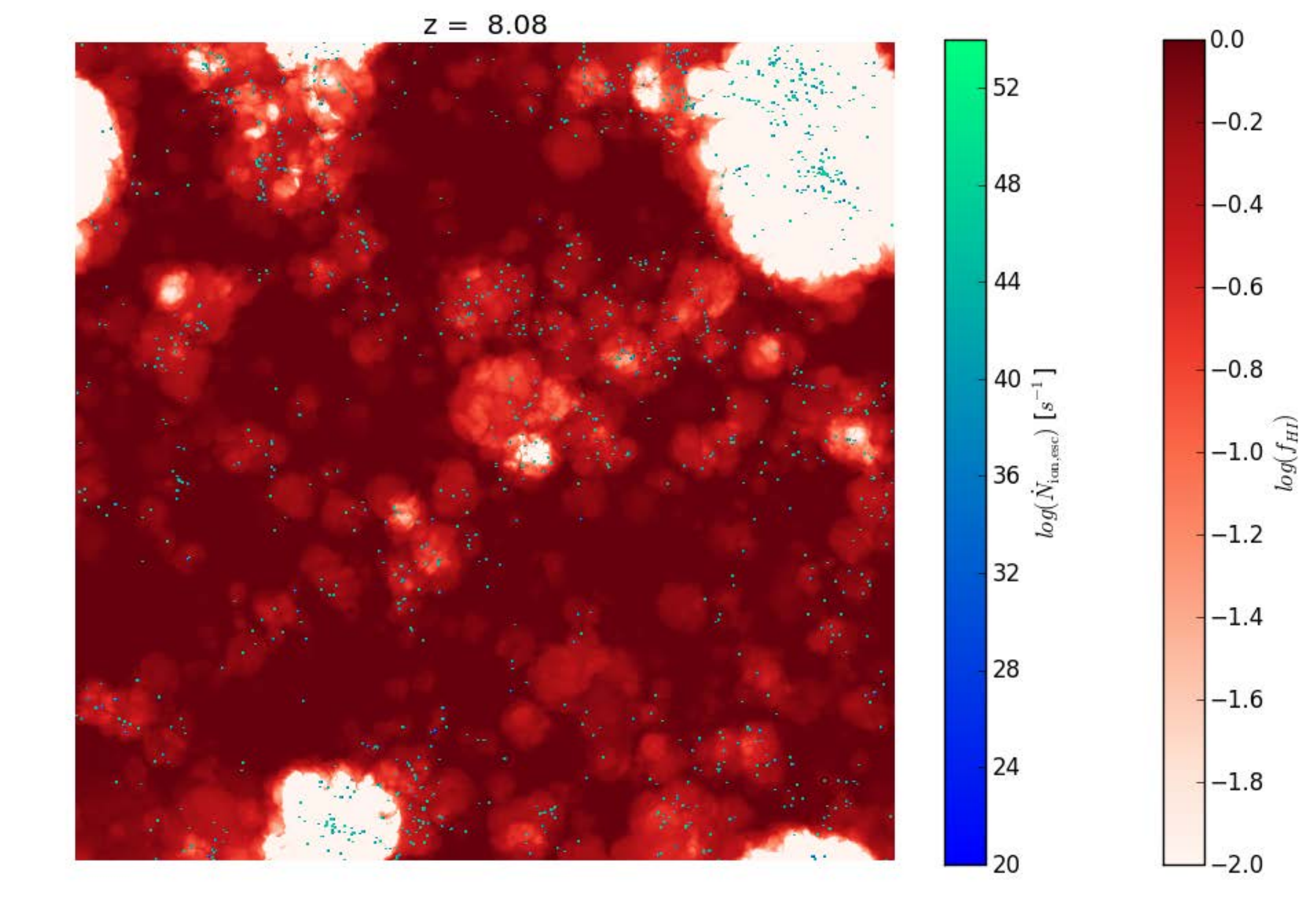}}}
\mbox{\includegraphics[width=0.95\columnwidth]{{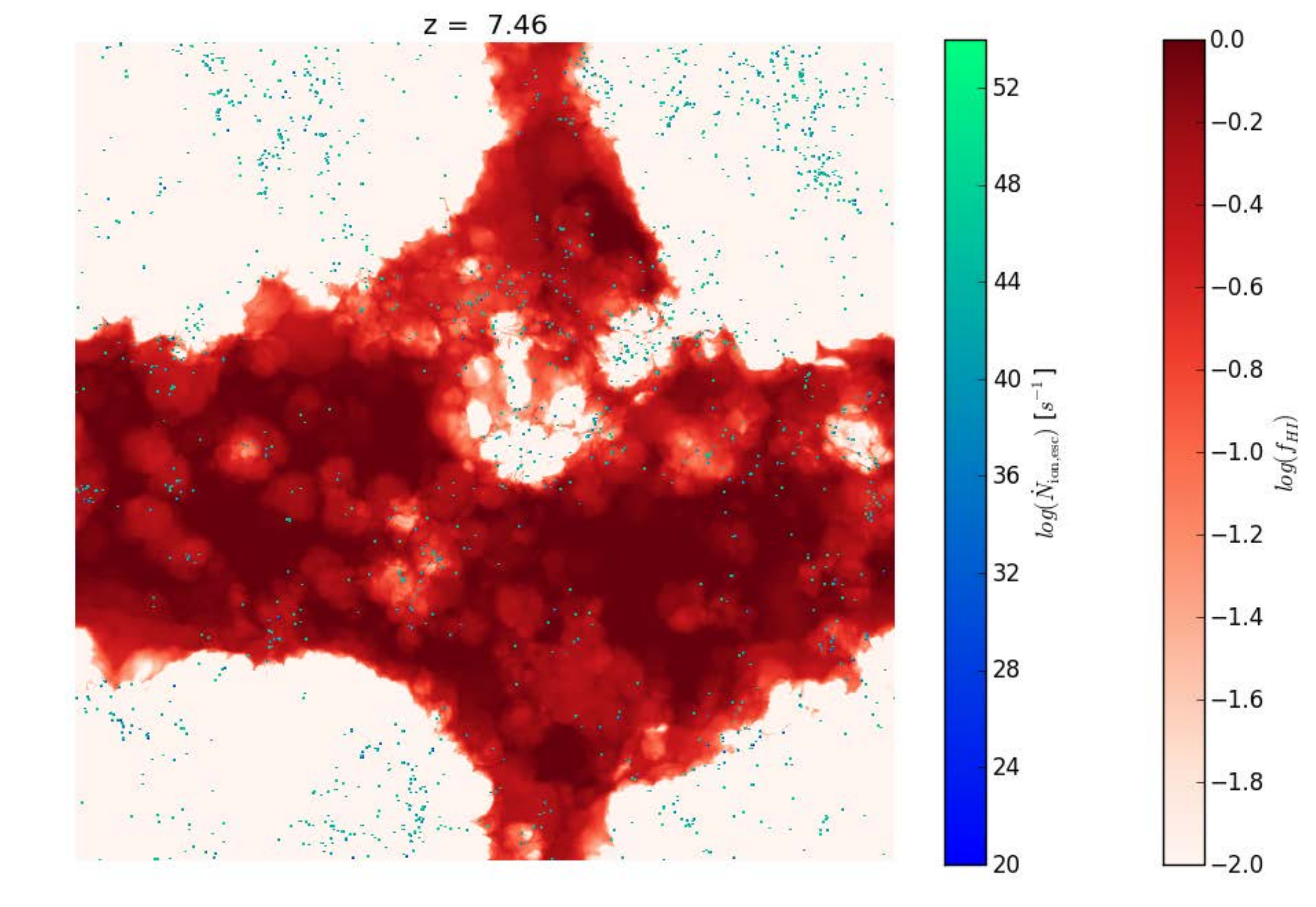}}}}
\centerline{
\mbox{\includegraphics[width=0.95\columnwidth]{{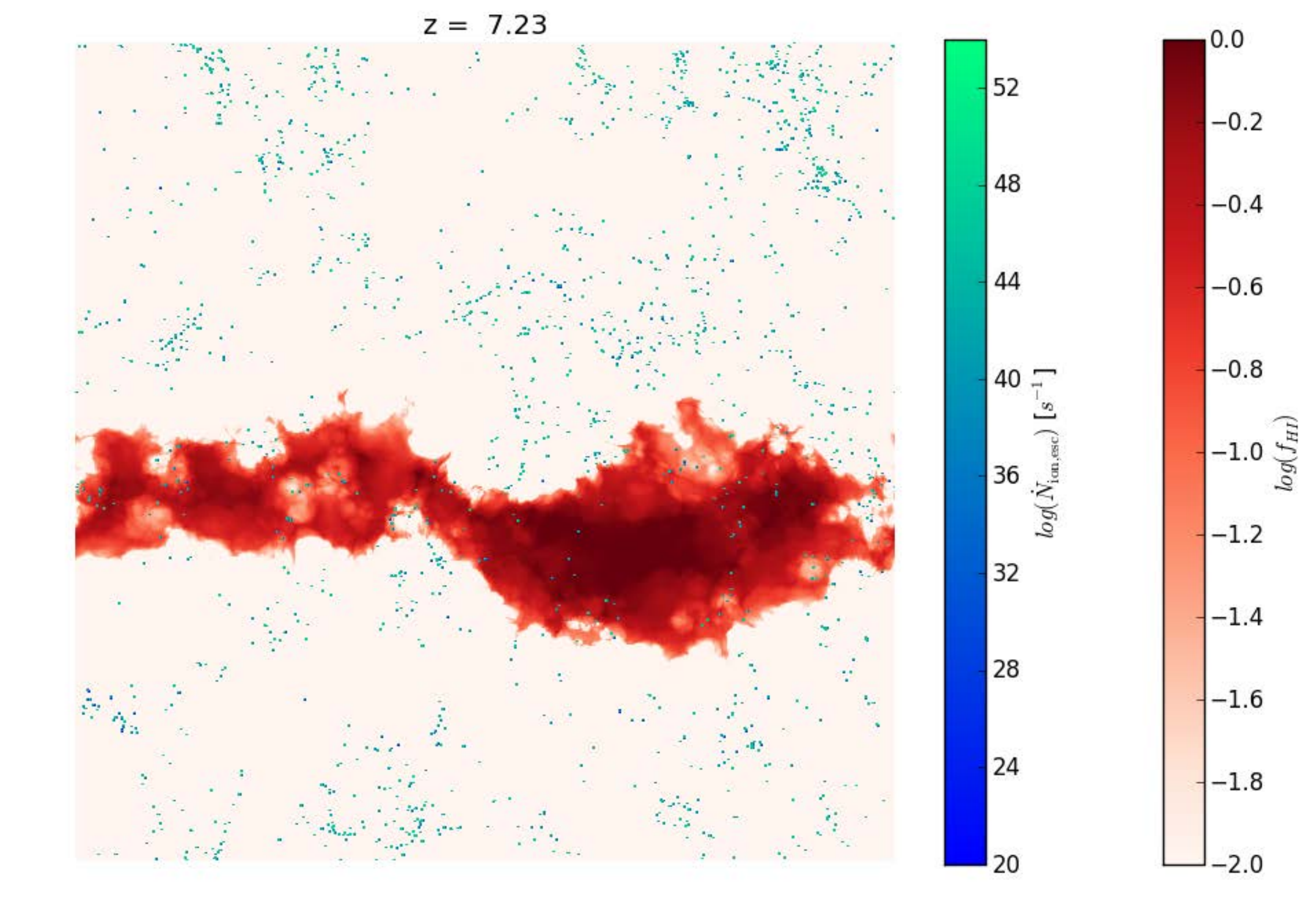}}}
\mbox{\includegraphics[width=0.95\columnwidth]{{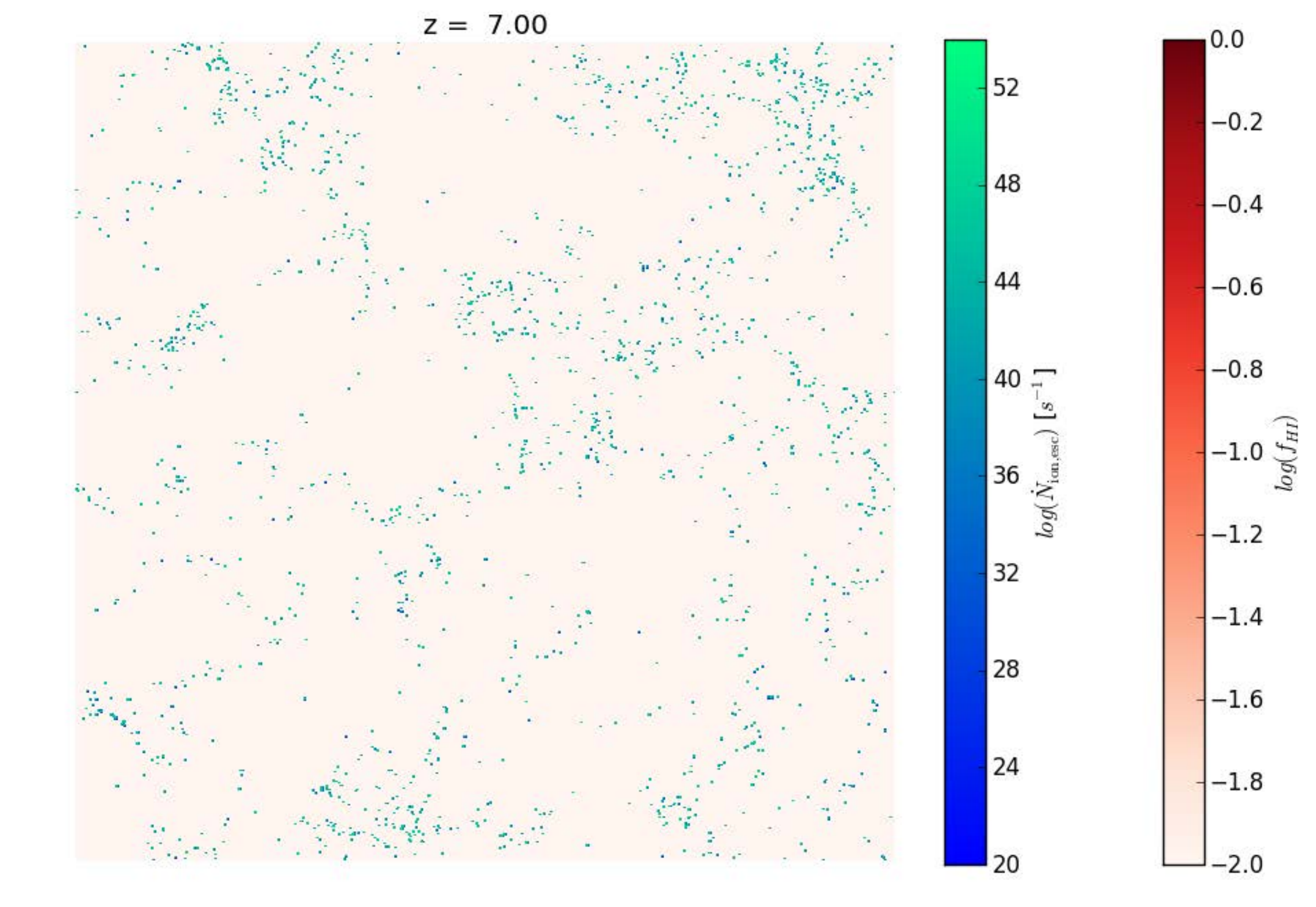}}}}
\end{center}
\caption[Projections of the logarithm of the neutral hydrogen fraction and ionizing emissivities]{Volume-weighted projections of the logarithm of the neutral hydrogen fraction and ionizing emissivities (colored pixels) through the 14.4 Mpc box at redshifts $z=11, 10, 9, 8, 7.46$ and $7$. See Fig. 7 caption for further explanation. This $1152^3$ simulation has identical mass and spatial resolution as the $512^3$ test simulation, but ionizes considerably earlier. }
\label{1152_cube}
\end{figure*}

Here we present the results of the science run carried out at identical mass and spatial resolution to the 512\_all run, but in a box 2.25 times the width. Because the volume is greater than $10 \times$ that of the former, we have much better statistical coverage of the ionizing sources at all redshifts, including more massive halos. 

Fig. \ref{1152_cube} shows how reionization proceeds through a series of projections of the \hi~ fraction through the width of the box. The color bar is chosen to accentuate the small \hii~ regions of low to moderate ionization fraction, while larger highly ionized \hii~ bubbles appear white. Superimposed as blue and green pixels is the ionizing emissivity field. One sees that before the HMACHs begin to dominate the total ionizing budget at $z \sim 8$, the volume is filled with small \hii~ regions which are only partially ionized ($f_i=0.01-0.1$). They increase in size and number, but are still largely isolated at $z=9$. By this time, a cluster of dozens of higher mass galaxies forms in the upper right hand corner of the box, and their combined ionizing flux drives a strong ionization front into the IGM. That it is clustered sources that drive the larger ionized bubble is particulary evident in the z=8.08 and 7.46 snapshots. Because of our small volume and periodic boundary conditions, this \hii~ superbubble fills the entire volume by $z=7.1$, sweeping over the smaller \hii~ regions as well as a smaller superbubble percolating in the center of the box. 

\begin{figure}
\includegraphics[width=0.9\columnwidth]{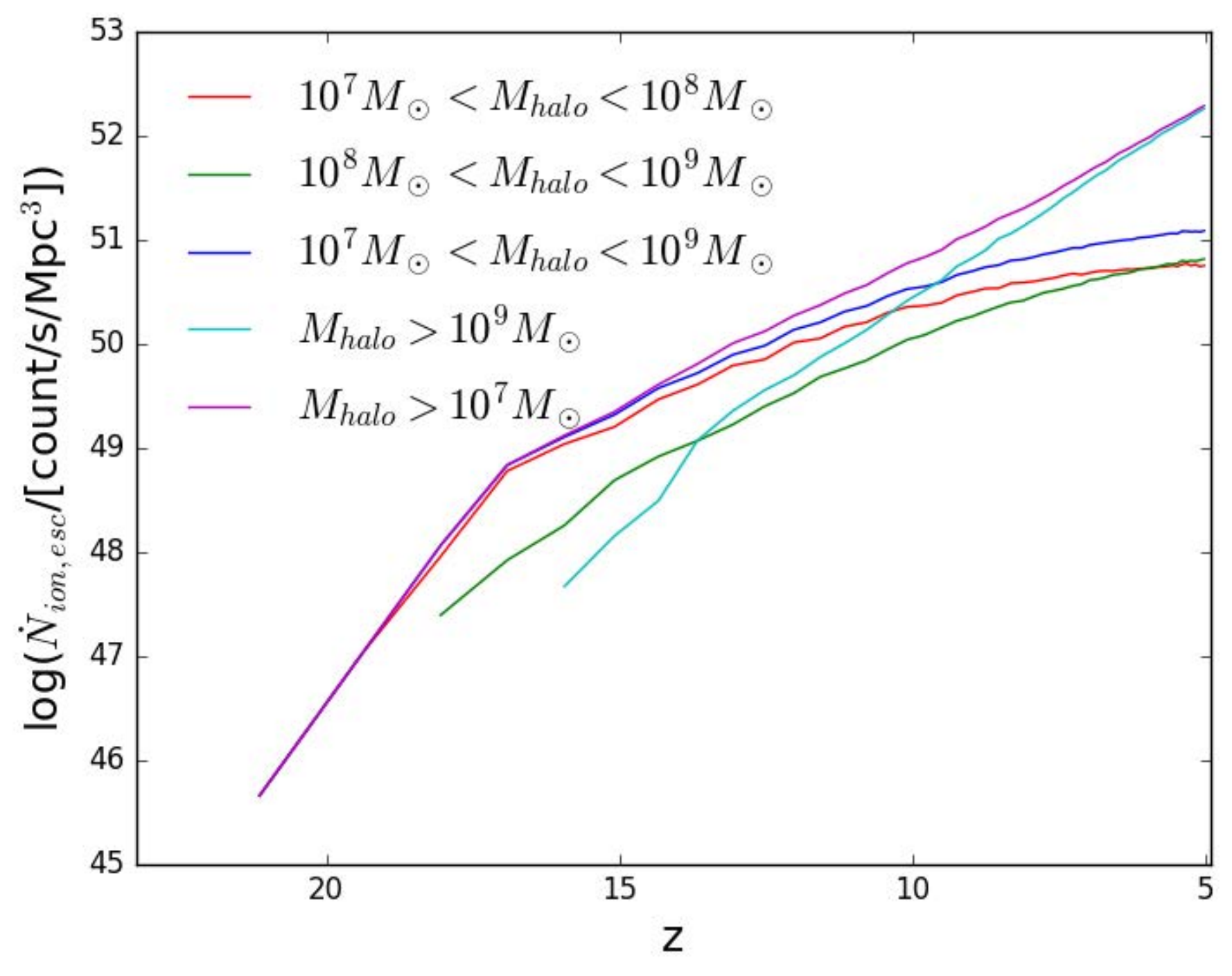}
\caption{Number of escaping ionizing photons coming from halos in different mass bins from the 1152\_all simulation. Red line: $10^7 \leq \mh/M_{\odot} \leq  10^8$; green line: $10^8 \leq \mh/M_{\odot}  \leq 10^9$; blue line: $10^7 \leq \mh/M_{\odot}  \leq 10^9$; turquoise line: $\mh/M_{\odot} > 10^9$. The purple line is the sum over all halos.}
\label{Ndot}
\end{figure}

Fig. \ref{Ndot} shows the redshift evolution of the number of ionizing photons escaping from halos in various mass ranges. The MCs (red line) begin contributing at $z \sim 22$ and dominate the LMACHs (green line) at all redshifts down to $z \sim 6$. This is due to their higher numbers and escape fractions as compared to the LMACHs. In fact the MCs dominate he HMACHs (turquoise line) until $z \sim 10$, and become subdominant thereafter. The total ionizing photon flux is shown by the purple line, and increases by three orders of magnitude from $\dot{N} \sim 10^{49}$ s$^{-1}$ Mpc$^{-3}$ at $z=17$ to $\dot{N} \sim 10^{52}$ s$^{-1}$ Mpc$^{-3}$ at $z=7$, when overlap occurs. 

\begin{figure}
\includegraphics[width=\columnwidth]{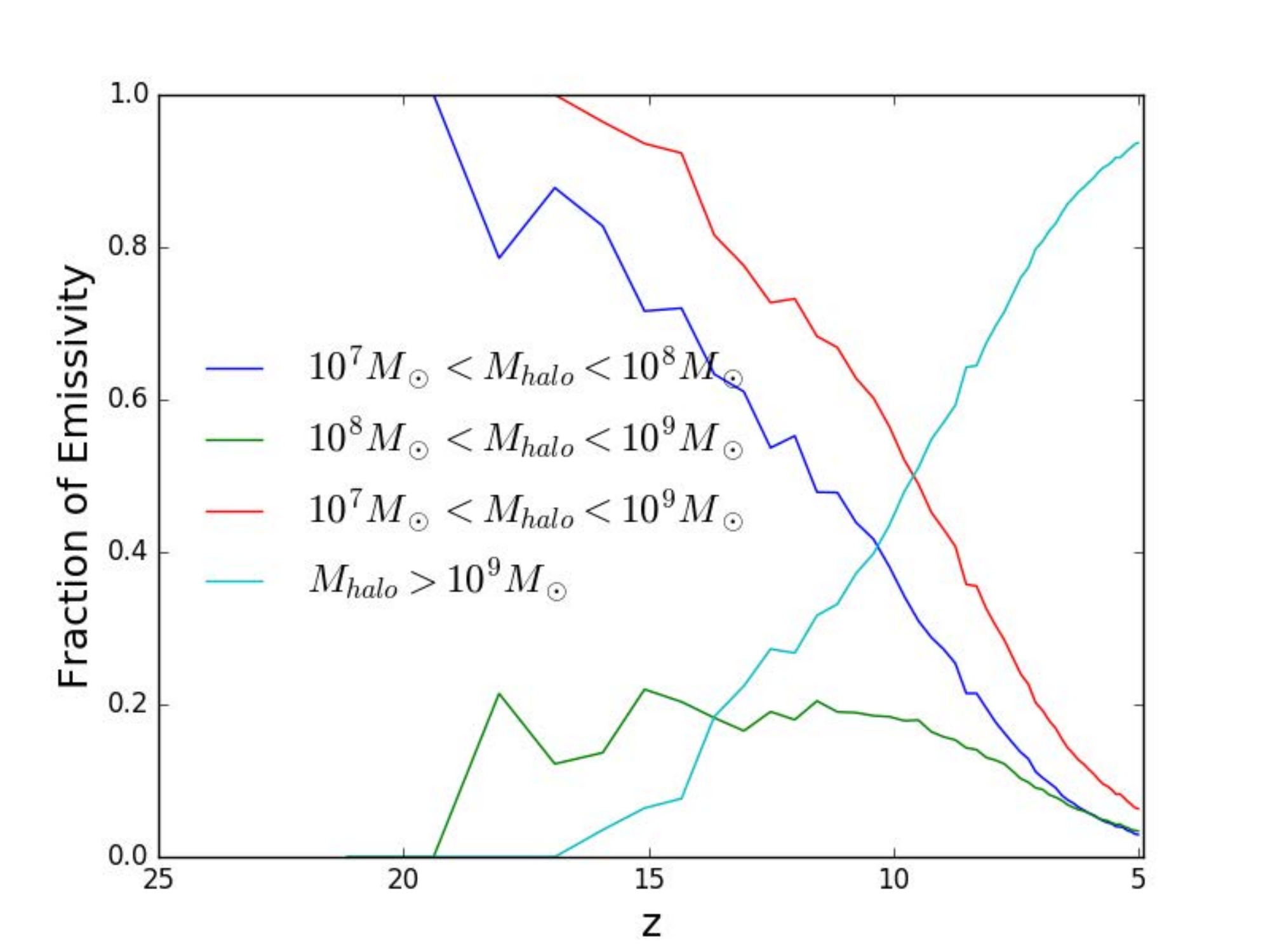}
\caption{The fraction of the total ionizing luminosity coming from halos in different mass bins from the 1152\_all simulation. Blue line: $10^7 \leq \mh/M_{\odot} \leq  10^8$; green line: $10^8 \leq \mh/M_{\odot}  \leq 10^9$; red line: $10^7 \leq \mh/M_{\odot}  \leq 10^9$; and turqoise line: $\mh/M_{\odot} > 10^9$. }
\label{Emis_frac_4}
\end{figure}

The relative contribution of the different halo mass bins is illustrated in Fig. \ref{Emis_frac_4}. As expected, MCs dominate the high redshift ionizing photon budget due to their high numbers and escape fractions. Interestingly, LMACHs are never more than a $\sim 20\%$ contibutor, due to their significantly lower escape fractions. The HMACHs begin forming at $z \sim 16$ in this simulation, and only begin to exceed the contribution of the MCs at $z \sim 10$, and of MCs+LMACHs at $z \sim 9$. This figure makes it clear that the contribution of the MCs to the early phases of reionization $15 \geq z \geq 10$ cannot be ignored, and is more significant than that of the LMACHs.

\begin{figure*}
\begin{center}
\centerline{
\mbox{\includegraphics[width=\columnwidth]{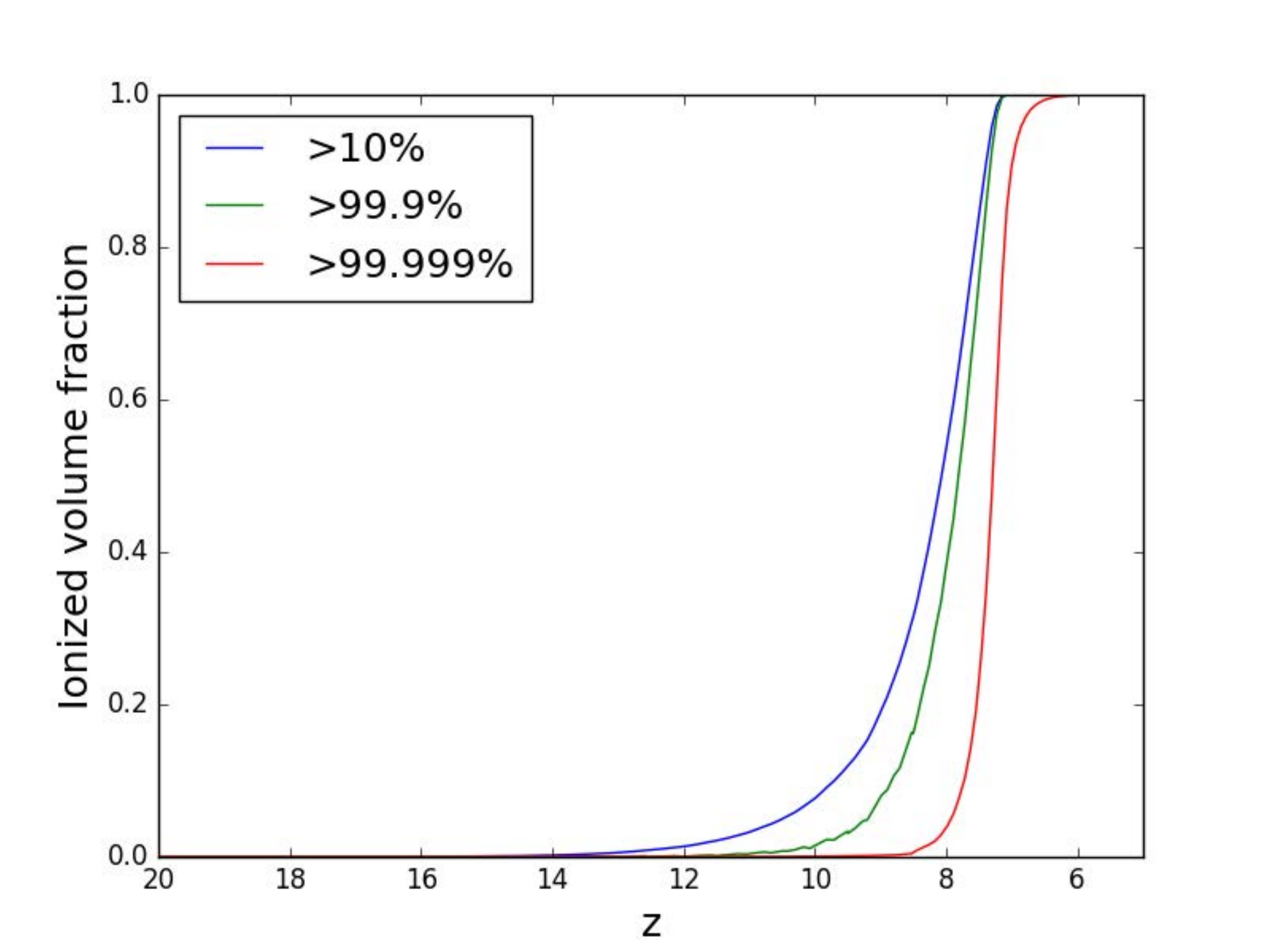}}
\mbox{\includegraphics[width=\columnwidth]{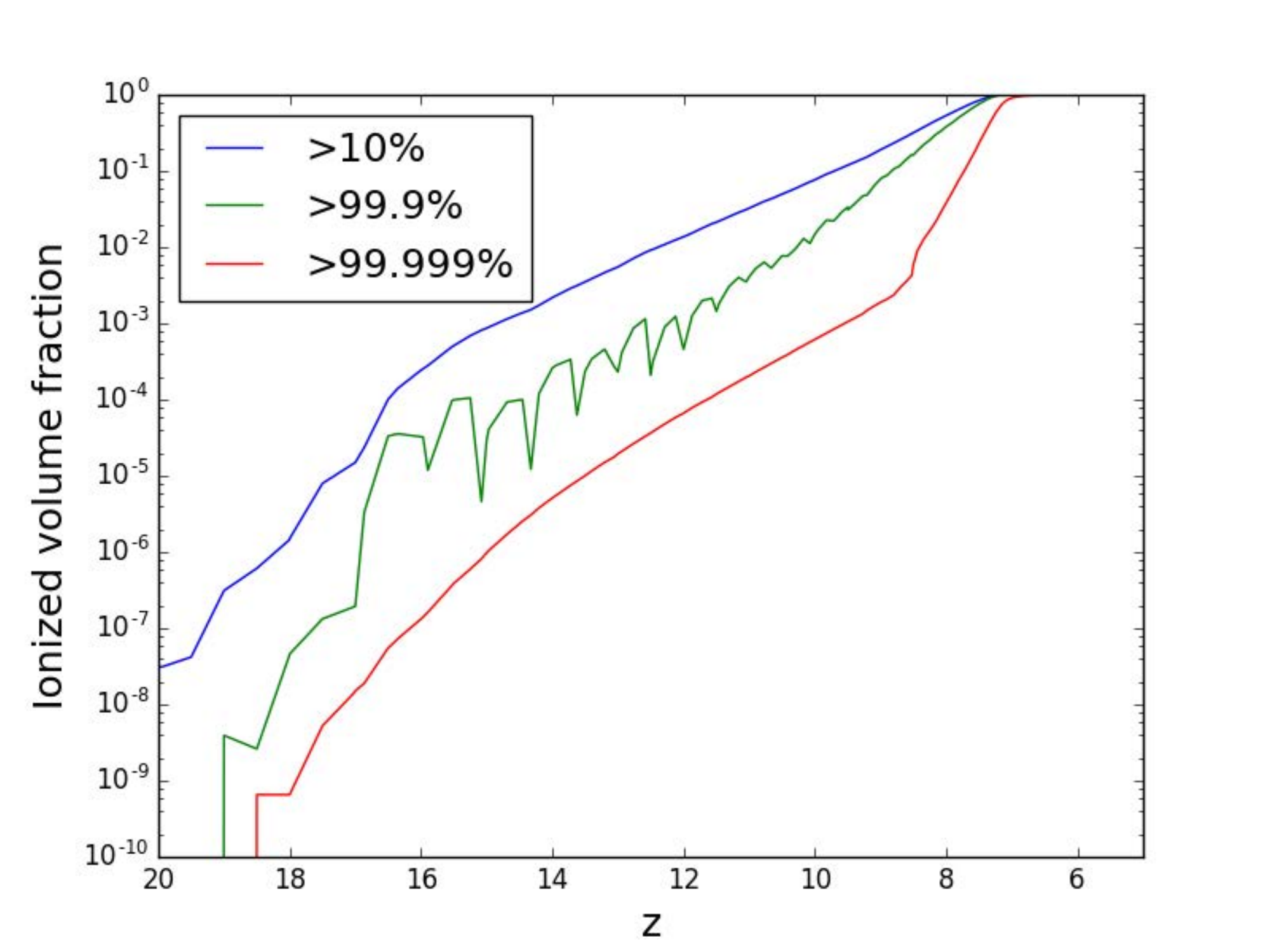}}}
\end{center}
\caption{Ionized volume fraction versus redshift for different ionization fraction thresholds. Top: linear scale, Bottom: logarithmic scale. Relic \hii~ regions show up prominantly in the 99.9\% curve as sources are reintroduced on 20 Myr intervals.   }
\label{ionized_volume}
\end{figure*}

Fig. \ref{ionized_volume} shows the evolution of the ionized volume fraction $f_i$ for three levels of ionization fraction \citep{So14}: 10\%, 99.9\%, and 99.999\%. The left(right) panel shows the linear(log) of the ionized volume fraction, respectively. Looking at the left panel first, we see that low levels of ionization (10\%) are obtained in larger fractions of the volume than high levels of ionization $\geq 99.9\%$ at all redshifts, but is more pronounced at high redshifts. As found by \cite{So14}, the curve for the highest level of ionization 99.999\% is significantly displaced to lower redshifts relative to the other two, and reaches $f_i = 1$ at $z=6$, a $\Delta z=1$ later.  

Looking at the right panel, we see by the blue curve that lower levels of ionization begin to occupy tiny fractions of the volume before $z=20$, consistent with the photon production history shown in Fig. \ref{Ndot}. The blue curve increases monotonically to lower redshifts, reaching $f_i = 1$ at $z=7.1$. The green curve shows the fraction of the volume that reaches the threshold of 99.9\% local ionization fraction. It is not monotonic, but shows a sawtooth like modulation. This is a consequence of our insertion of a new set of ionizing sources every 20 Myr. While the periodicity is an artifact of our insertion method, some variability in $f_i$ would be expected at early times in the continuous insertion limit as star formation in low mass halos turns on and off, creating relic \hii~ regions in the process. 

\section{Discussion and Conclusions}
\label{sec:discussion}

In this paper, we study the role of low mass halos in reionization. Using fully-coupled cosmological radiation hydrodynamic simulations with a subgrid model for the ionizing sources derived from the {\it Renaissance Simulations} \citep{Xu16}, we find that galaxies of dynamical mass $10^7M_\odot<\mh<10^8M_\odot$ make an important contribution to the earliest stages of reionization, in agreement with the findings of \cite{Wise14}. Halos in this mass range have been neglected in previous studies because they were assumed to form stars at negligible rates due to inefficient $H_2$ cooling. However \cite{Wise14} showed that these so-called metal cooling halos (MCs) can cool by metal fine-structure lines and form stars if they have been enriched by ejecta from Pop III supernovae. Although the star formation efficiency is still quite low, this is compensated for by the MCs high space density and ionizing escape fraction. Interestingly, we find that MC's ionizing contribution dominates that of the LMACHs ($10^8M_\odot<\mh<10^9M_\odot$) over the entire redshift range before HMACHs ($\mh>10^9M_\odot$) become the dominant ionizers at $z \approx 10$. 

\begin{figure*}
\centering \includegraphics[width=1.7\columnwidth]{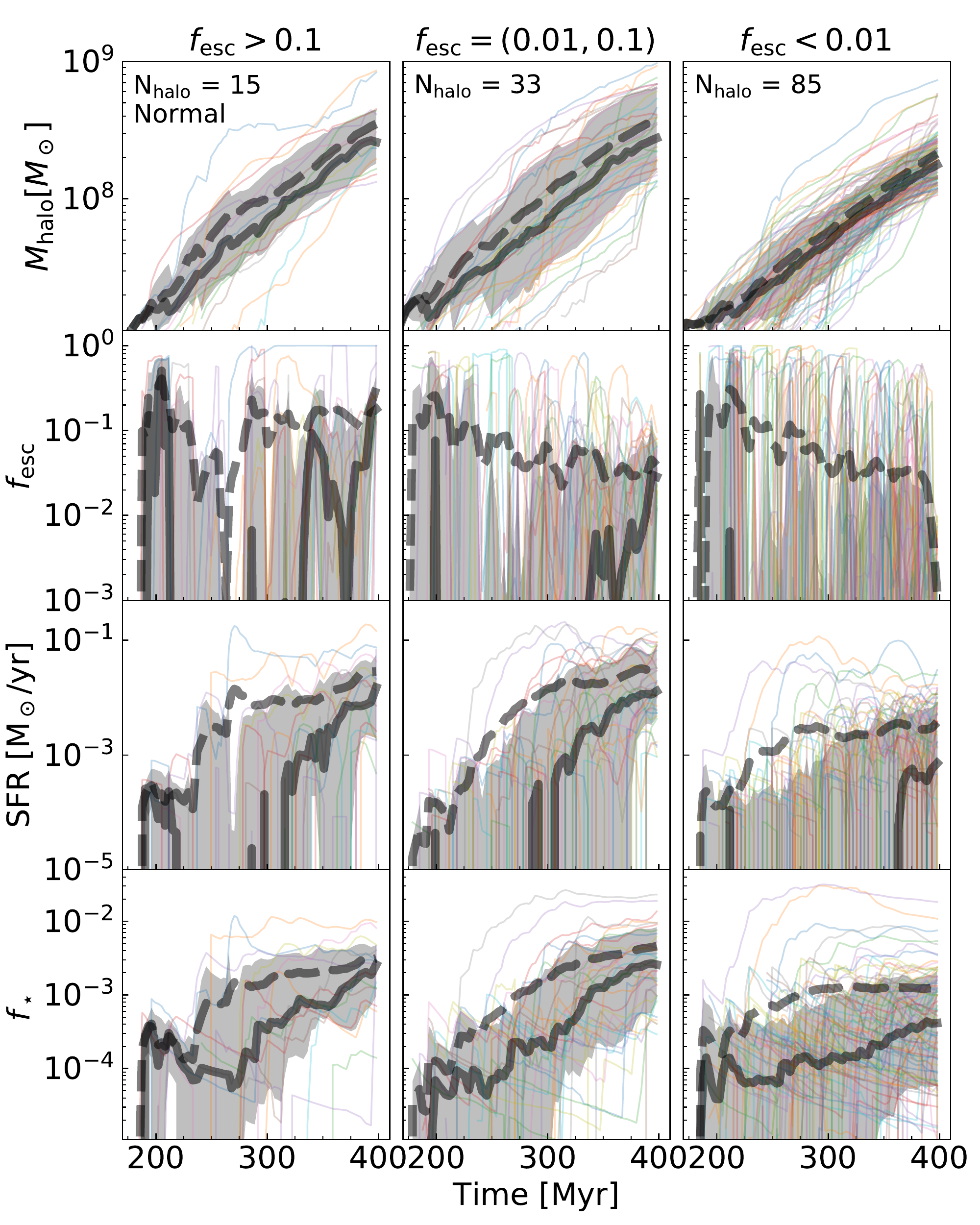}
\caption{Evolution of the LMACH population in the Normal simulation. The LMACHs are divided into three categories depending on their $f_{esc}$ at z=12.5: high ($f_{esc}>0.1$, left column); mid ($f_{esc}=(0.01,0.1)$, center column); low ($f_{esc}<0.01$, right column). 
The number of halos in each category are given in the upper left corner of the first row of figures.
Going down each column we plot the halo virial mass $M_h$, the ionizing escape fraction $f_{esc}$, the star formation rate (SFR), the stellar and gas fractions $f_{*}=M_{*}/M_h, f_{gas}=M_{gas}/M_h$, and the ratio of the total ionizing emissvity to the number of HI atoms within the virial radius. Thin lines plot individual halo trajectories. The thick dashed line plots the mean of the distribution at each time. The thick line and shaded regions shows the median and 33\% and 68\% percentiles.} 
\label{fig:fesc-lmach}
\end{figure*}

\subsection{Contribution of LMACHs to reionization}
\label{subsec:lmachs}
The low contribution of LMACHs to reionization is a surprising result of this study, one that bears some discussion. It is a consequence of the low ionizing escape fractions documented in Figs. 18-20 of \cite{Xu16}. These consistently show median values of less than 5\% for all simulations and redshifts in the LMACH mass range. Only for LMACHs near the top of the mass range do we see escape fractions of 10\% or more. To understand the physical origin of the low escape fractions we performed an analysis of the history of all 133 LMACH halos in the Normal Renaissance Simulation. We divided these into three groups according to their ionizing escape fraction at z=12.5: ``High", $f_{esc}>0.1$ (15 halos); ``Mid", $f_{esc}=(0.01,0.1)$ (33 halos); and ``Low",  $f_{esc}<0.01$ (85 halos). The results are shown in Fig. \ref{fig:fesc-lmach}. The columns show the 3 samples, while the rows show, respectively, the time evolution of halo mass, $f_{esc}$, star formation rate (SFR), the stellar mass fraction $f_{*}=M_{*}/M_h$, and gas mass fraction $f_{gas}=M_{gas}/M_h$ over the roughly 225 Myr interval preceeding z=12.5. Each panel plots the trajectories of individual halos as thin colored lines and their mean as a thick dashed line.  The thick line and shaded regions shows the median and 33\% and 68\% percentiles.

Looking at the $f_{esc}$ progression, one can see that the High halos have a mean value that's above 0.1 about 75\% of the time, whereas the Mid and Low halos steadily decline below 0.1 with the Low halos'  $f_{esc}$ values plummeting in the last 50 Myr.  What causes this dramatic drop in the Low sample? To explain this we note that $f_{esc}$ histories of individual halos in all three samples show that $f_{esc}$ is stochastic having large values for 25-50 Myr, coming from the bursts of star formation.  The $f_{esc}$ values are either close to unity or << 1\%.  We believe that the $f_{esc}$ < 1\% filter which defines the Low sample is selecting halos in their quiescent phases, where they have some active star formation, but the photons aren't escaping at z=12.5 for various reasons. Possible reasons are: (1) the young stars are not luminous enough; (2)  stars just recently formed and the \hii~ region is still trapped within the halo; or (3) active star formation is ending and the halo is recombining. According to this explanation, the Mid and High samples were simply caught in a different phase of the stochastic star formation process. The mean value of 5\% for the total LMACH sample is therefore the result of averaging many Low halos with with a smaller number of Mid and High halos with higher escape fractions at the instant when the samples were defined. These average to $f_{esc}(LMACH) \approx$ 5\%. This value could also be viewed as a temporal average of $f_{esc}$ in an individual LMACH halo. 

\subsection{Limitations}
\label{subsec:limitations}
Our simulations suffer from several limitations, which we discuss here. The first is the omission of the contribution of Pop III stars to the ionizing photon budget. Pop III star formation is not included in our simulation, and therefore, the very first sources of ionizing photons are omitted entirely. Simulations that include this physics show that the Pop III SFR increases for the first 100 Myr, and then remains approximately constant at a level of $5 \times 10^{-5}$ M$_{\odot}$ yr$^{-1}$ Mpc$^{-3}$ until overlap \citep{Wise12_Galaxy,Xu16a}. Assuming an ionizing photon to stellar baryon number ratio of 60,000 \citep{Schaerer02}, then this corresponds to a constant ionizing flux from Pop III stars of $1.1 \times 10^{50}$ s$^{-1}$ Mpc$^{-3}$.  Referring to Fig. 6 we see that small galaxies begin to exceed this average emissivity at $z \approx 13$ and dominate it by 1.5 orders of magnitude by overlap. \cite{Wise14} presented one-zone reionization models with and without Pop III stars, and found that their inclusion boosted the ionized mass fraction by less than 10\% at all redshifts $z<20$, and increased $z_{ov}$ by a small amount. We therefore conclude that the results presented here would not be qualitatively changed by the inclusion of Pop III stars.  

The second limitation is the use of a non-evolving mapping from halo mass to escaping ionizing emissivity. As described previously, Table 1 is taken from an analysis of a $z=12.5$ snapshot of the Normal Renaissance Simulation. Ideally, one would like to use a redshift-dependent set of tables that span the entire reionization history. One can ask how different might the results be if we had done that? To address this we rely on the extensive discussion of the evolution of the ionizing escape fraction in the {\it Renaissance Simulations} in \cite{Xu16}, Sec. 4. They found that $f_{esc}$ depends primarily on halo mass, and is rather insensitive to redshift or large-scale environment. This is illustrated in Figs. 18-20 of \cite{Xu16}, which shows $f_{esc}$ and $\dot{N}_{ion,esc}$ versus $\mh$ for several redshifts from each of the three simulations. The figures show that evolution of the ionizing photon budget comes almost entirely from the evolving halo population, and not from the $f_{esc}$--$\mh$ relation which is remarkably constant with respect to redshift and environment. This is true for both the median as well as the scatter, which increases significantly below $10^8\Ms$. This is one input to our emissivity model. The other input is the probability that a halo of a given mass is actively forming stars at a given time. \cite{Xu16} show that this probability declines from unity at $\mh \approx 10^{8.5}\Ms$, is roughly 50\% (10\%) at $\mh =10^{8} (10^7)\Ms$, and zero below $\mh \approx 10^{6.5}\Ms$. However, for a given region these fractions decrease with time for halos below $10^8\Ms$ because of the increasing effects of radiative feedback from more massive galaxies which increase the filtering halo mass for efficient cooling and star formation. This effect is not included in our model. During the onset of reionization, \hii~ regions are primarily isolated before the overlap phase, and these galaxies are regulated from their own radiative and supernova feedback, unaffected from radiation originating from other galaxies. Therefore, the results from the Renaissance Simulations provides a good estimate of the ionizing photon production from these early galaxies. Only afterwards does the filtering effect for MCs becomes important precisely when the HMACHs begin to dominate the ionizing budget at $z \approx 10$. Therefore our main conclusion that MCs make an important contribution to early reionization, but have minimal contribution to the late stages of reionization, is unaltered by omitting the effect of filtering.

The third limitation also arises from our choice of the z=12.5 state of the Normal simulation. It has a distribution of escaping ionizing luminosities with a median that is an order of magnitude lower than the combined (RP+N+V)/3 sample in three of the four bins in the LMACH mass range, previously shown in Figure \ref{fig:halo-counts}.  If we assume that the combined sample is the more accurate one, we can estimate how many ionizing photons are being neglected, using the absolute and fractional emissivities shown in Figures \ref{Ndot} and \ref{Emis_frac_4}, respectively.  The LMACHs produce a factor of three less escaping ionizing photons than the MCs at $z > 10$.  If we have underestimated this factor by ten (an upper limit, given that one of the four mass bins is not deficient), the total emissivity would be boosted by a factor $(\dot{N}_{\rm MC} + \dot{N}_{\rm LMACH,adj}) / (\dot{N}_{\rm MC} + \dot{N}_{\rm LMACH}) \sim 3$, taking the boosted emissivity $\dot{N}_{\rm LMACH,adj} = 10\dot{N}_{\rm LMACH}$ and $\dot{N}_{\rm LMACH} \sim \dot{N}_{\rm MC}/3$.  In this extreme limit, LMACHs would no longer be subdominant to the MCs, as depicted in Fig. 5. These additional photons would result in a slightly higher ionized fraction at $z > 10$ when MCs and LMACHs dominate the ionizing photon budget and when the ionized volume fraction is only a few percent (Figure \ref{ionized_volume}).  However at later times when larger halos dominate, this difference would become unimportant, minimally impacting the late stages of reionization along a similar argument in the previous limitation.

A fourth limitation is that our simulation volume is not large enough to adequately represent the characteristics of reionization on the largest scales. Our requirement to include halos as small as $10^7\Ms$ dictated the choice of a small box size to keep the computational cost reasonable.  According to \cite{Iliev14}, a comoving volume of $\sim$100 Mpc/h per side is needed for simulating a convergent mean reionization history. We therefore cannot claim our $z_{ov}$ is converged. Running a 100 Mpc/h box while resolving $10^7\Ms$ halos is not feasible on current supercomputers. However one could adopt the approach of \cite{Ahn12}, who incorporated the contribution of Pop III ionizing sources in a large scale reionization simulation via a subgrid model derived from a small box simulation. The approach would carry over to the case of MCs, with some modifications needed to take into account the stochastic nature of star formation is such halos. 

\subsection{Comparable works}
\label{subsec:works}
As this manuscript was being finalized for submission, a new paper appeared which directly addresses the same question we are exploring: the role of the smallest galaxies to reionization. 
\cite{Kimm17} perform AMR simulations of high redshift galaxies very similar in design and objectives as \cite{Wise14} and found very similar results. Specifically, they used the RAMSES-RT code \citep{Rosdahl15} to simulate the assembly of the first galaxies in a 2 comoving Mpc box including the feedback by Pop III stars forming in minihalos. Importantly, they cover the halo mass range $10^7 \leq M/M_{\odot} \leq 10^8$, albeit with smaller samples than \cite{Xu16}. While the numerical methods differ from those employed by \cite{Wise14} and \cite{Xu16}, the physical ingredients are essentially the same, including details of the star formation and feedback models. Their results can thus be compared directly. 

\cite{Kimm17} find that the mean ionizing escape fraction increases with decreasing halo mass, in agreement with \citep{Wise14,Xu16}, although they quote lower values of 20-40\% for the smallest halos. They find that star formation in individual halos is stochastic, with recovery times from stellar feedback ranging from $\sim$ 20 to 200 Myr. They do not compute a halo occupation fraction of active star formation versus halo mass as done by \cite{Xu16}. However the $M_{\rm star}$ vs. $\mh$ data shown in their Fig. 5 is in good agreement with \citet{Xu16}'s results for $\mh > 10^7\Ms$, but displays much lower stellar masses for smaller halos.

\cite{Kimm17} do not perform an inhomogeneous reionization simulation as we do, but they do present a one-zone model using their results and those from \cite{KimmCen14} as input from which they derive conclusions about the importance of minihalos ($\mh < 10^8\Ms$) to reionization. Here we compare our conclusions and highlight points of agreement and uncertainty. We both agree that minihalos dominate the earliest stages of reionization and that HMACHs dominate the late stages of reionization leading to overlap. The redshift at which the HMACHs become dominant is similar in both models: $z\approx 11$ in \cite{Kimm17} and $z\approx 10$ in this work. Where we differ is the relative contribution of LMACHs and minihalos (which includes what we are calling MCs). \cite{Kimm17} find that LMACHs begin to dominate minihalos at $z\approx 17$ (see their Fig. 15), whereas we find they are subdominant at all redshifts prior to overlap (Fig. 7). However, as we discussed in Sec. \ref{subsec:limitations}, we may have underestimated the LMACH contribution by as much as a factor of 10 due to our choice of ionizing emissivities from the z=12.5 Normal simulation rather than use an average of the (V+N+RP) sample. On the other hand, the contribution of LMACHs in the mass range $10^8 \leq \mh/\Ms \leq 10^{8.6}$ in \cite{Kimm17} is based on an extrapolation from data presented in \cite{KimmCen14} of more massive halos. The origin of this discrepancy would seem to be the lower stellar masses found by \cite{Kimm17} as compared to \cite{Xu16} for halo masses below $\mh/\Ms \approx 10^7$, their lower ionizing escape fractions in MC halos, and their assumption of higher ionizing escape fractions for LMACHs in their fiducial model compared to \cite{Xu16}. These two effects would tilt the contribution of ionizing photons away from MCs and toward LMACHs. It is important to note that \cite{Kimm17} have only 6 galaxies in the LMACH mass range while we have 133. Their conclusions on the relative importance of MCs relative to LMACHs to reionization is based on the above-mentioned  extrapolation.

\cite{Kimm17} also consider a ``Low" model which adopts low values of $f_{\rm esc}$ for LMACHs, consistent with \cite{Xu16}. But the Low model has a constantly declining $f_{\rm esc}$ as a function of $\mh$, which considerably underpredicts $f_{\rm esc}$ for HMACHs when compared to our simulations.This model reionizes late and underpredicts $\tau_{\rm es}$ by a considerable factor.  Given the uncertainties discussed above, we believe it is premature to conclude that minihalos are unimportant for reionization. Their contribution may be on the same order as LMACHs above $z=10$. However, we agree with \cite{Kimm17}'s conclusion that halos as small as $10^8\Ms$ must be included in reionization simulations. Our research only highlights the need to better understand the properties of LMACH galaxies through more comprehensive simulations. We have carried out a detailed analysis of the nearly 400 LMACHs galaxies in the combined (V+N+RP) sample of the Renaissance Simulations and will report on the results of this analysis in a future paper.

\acknowledgments

We wish to acknowledge the anonymous referee whose questions and comments improved the content and quality of the manuscript. This research was supported by National Science Foundation (NSF) grants
AST-1109243 and AST-1615848 to M.L.N. J.H.W. acknowledges support from NSF grant AST-1333360 and Hubble Theory grants HST-AR-13895 and HST-AR-14326. The simulations were performed using the ENZO code on the Comet  system at the San Diego Supercomputer Center, University of California San Diego with support from XRAC allocation MCA-TG98020N to M.L.N. This research has made use of NASA's Astrophysics Data System Bibliographic Services. The majority of the analysis and plots were done with YT \citep{yt_full_paper}. ENZO and YT are developed by a large number of independent researchers from numerous institutions around the world. Their commitment to open science has helped make this work possible. 
\bibliography{ms}
\bibliographystyle{apj}

\end{document}